\documentclass[prb,twocolumn,showpacs,preprintnumbers,superscriptaddress,amsmath,amssymb]{revtex4}

\usepackage{graphicx}
\usepackage{dcolumn}
\usepackage{bm}
\usepackage{color}

\begin{document}

\title{A study on correlation effects in 
two dimensional topological insulators
}

\author{Y. Tada}
\affiliation{Institute for Solid State Physics, The University of Tokyo,
Chiba 277-8581, Japan}
\author{R. Peters}
\affiliation{Department of Physics, Kyoto University, Kyoto 606-8502, Japan}
\author{M. Oshikawa}
\affiliation{Institute for Solid State Physics, The University of Tokyo,
Chiba 277-8581, Japan}
\author{A. Koga}
\affiliation{Department of Physics, Tokyo Institute of Technology, 
Tokyo 152-8551, Japan}
\author{N. Kawakami}
\affiliation{Department of Physics, Kyoto University, Kyoto 606-8502, Japan}
\author{S. Fujimoto}
\affiliation{Department of Physics, Kyoto University, Kyoto 606-8502, Japan}

\newcommand{\vecc}[1]{\mbox{\boldmath $#1$}}

\begin{abstract}
We investigate correlation effects in two dimensional
  topological insulators (TI).
In the first part, we discuss finite size effects 
for interacting systems of different sizes in a ribbon geometry.
For large systems, 
there are two pairs of well separated massless modes
on both edges. For these systems, we analyze the finite size effects
using a standard bosonization approach.  
For small systems, where
the edge states are massive Dirac fermions, we use the inhomogeneous
dynamical mean field theory 
(DMFT) combined with iterative perturbation theory as an impurity
solver to study interaction effects.
We show that the finite size gap
in the edge states is renormalized for weak interactions, 
which is consistent with a Fermi-liquid picture for small size TIs.
In the second part, 
we investigate phase transitions in finite size TIs at zero temperature
focusing on the effects of possible inter-edge Umklapp scattering for the
edge states within the inhomogeneous DMFT using
the numerical renormalization group.
We show that correlation effects are effectively stronger near the edge
sites because the coordination number is smaller than
in the bulk.
Therefore, the
localization of the edge states around the edge sites,
which is a fundamental property in TIs, is weakened for strong coupling
strengths.  
However, we find no signs for "edge Mott insulating states" and
the system stays in the topological insulating state,
which is adiabatically connected to the non-interacting state,
for all interaction strengths smaller than the critical value.
Increasing the interaction further, a
nearly homogeneous Mott insulating state is stabilized.
\end{abstract}

\pacs{Valid PACS appear here}

\maketitle
\section{Introduction}
\label{sec:intro}

Topological insulators (TI) and topological superconductors (TSC) have been
attracting great interest since the theoretical studies on
the quantum spin Hall effect, 
\cite{KM05_1, KM05_2, BHZ06, Bernevig06, Murakami06}
and their experimental realization.\cite{Konig07, Roth09, Hsieh08,
  Hsieh09}  
They can be characterized by an energy gap in the bulk and gapless
edge or surface 
states at their boundaries.\cite{Hasan10,Qi11}
The direct relation between the bulk gap and the edge states, which is
called the bulk-edge correspondence, is considered
to be one of the most important properties in TIs and TSCs.
These edge states are protected by topological properties of the
bulk system, and are therefore robust against perturbations 
such as disorder or interactions
which do not break the symmetries of the system.\cite{Wu06,Xu06}
This suggests that, as long as the bulk remains gapful with
non-trivial topology, the edge states exist at its boundaries.
Correspondingly, in non-interacting systems, 
topological quantum phase transitions between topologically 
trivial and non-trivial phases require that
the bulk gap is continuously closed.
These gapped non-interacting systems which support the bulk-edge
correspondence can be classified by ten distinct classes
depending on their symmetries.\cite{Schnyder08, Schnyder09,Kitaev09}

However, it is not clear to what extent
the bulk-edge correspondence holds in the presence of
interactions.
Interactions are expected to create novel 
correlated topological states,\cite{Pesin10,Gurarie11,Rachel10,Hohenadler10,Hohenadler11_1,Hohenadler11_2,Yamaji11,Yu11,Yoshida11,
Fidkowski10_1,Fidkowski10_2,Turner11}
such as TIs/TSCs ${\it without}$ edge or surface states. 
For example, it has been proposed that strong correlations could 
realize a topologically non-trivial state without edge states
when interactions become large.\cite{Gurarie11}
In this scenario, additional ``zeros'' in the Green's functions
develop inside the bulk gap 
and finally merge with the poles which correspond to the edge states.
Thus, the resulting state does not possess any edge states although it
remains gapped in the bulk. 
Among the models exhibiting topological insulating states,
the Kane-Mele-Hubbard models have been extensively studied and
the obtained phase diagrams show that the TI states extend to
a wide region.\cite{Rachel10,Hohenadler10,Hohenadler11_1,Hohenadler11_2,Yamaji11,Yu11}
For large values of the Hubbard interaction $U$,
magnetic states related to the honeycomb lattice structure
are stabilized. 
In between, a possible realization of
topological states without edge states has been proposed.\cite{Yamaji11}
Correlation effects have also been investigated in one dimensional 
systems.\cite{Fidkowski10_1,Fidkowski10_2,Turner11}
It has been shown that, in time-reversal symmetric TSCs
with inter-chain interactions,
the ground states are characterized by fermion parity in addition
to the time-reversal symmetry.
These symmetries can distinguish different topological states
and the ground states 
are labeled by topological numbers in $\mathbb{Z}_8$ instead of
$\mathbb{Z}$ for non-interacting systems.
Furthermore, it was pointed out that in some cases the edge states actually
disappear as interactions are increased without any gap closing.

Besides the possible realization of novel states in 
TIs and TSCs, strong interactions affect their
quantitative properties. 
This is also the case for finite size systems, where the
topological character would be smeared.
Actually, it was pointed out that the edge states
can be gapped due to tunneling processes between
the two sides of the sample, and physical properties
strongly depend on external parameters
such as temperature and magnetic fields.\cite{Zhou08, Linder09, Lu10, Wada11}
Finite size effects have been investigated 
also in the context of dimensional crossovers.
\cite{Liu10,Potter10,Zhou11,YZhang10,Sakamoto10}
However, finite size effects in the presence of
interactions are still not well understood, and
properties of finite size TIs and TSCs
would be sensitive to interactions
in addition to external parameters.
Moreover, correlation effects between the two sides of the system
could have significant scattering processes
which might lead to novel inter-edge correlated states.\cite{Tanaka09,Hou09,Strom09,Teo09}

In this article we discuss correlation effects in 
two dimensional topological insulators in a ribbon geometry
at zero temperature.
In the first part, we discuss finite size effects 
in interacting TIs which generate
a gap in the edge states.\cite{Zhou08, Linder09, Lu10, Wada11,
  Liu10,Potter10,Zhou11,YZhang10,Sakamoto10} 
For large systems with $L_y\gg \xi_{\rm TI}$, 
where $L_y$ is a width of the system and $\xi_{\rm TI}$
is a characteristic localization length of the edge states,
the massless edge states at each side
are well defined and can be described by
Tomonaga-Luttinger liquids (TLL).
\cite{Tanaka09,Hou09,Strom09,Teo09}
We study effects of electron tunneling
between the two edges by
the standard bosonization method, and evaluate the finite
size gap in spin and charge sectors.

Next, we concentrate on relatively small systems, for which the edge
states are massive Dirac fermions.
For this purpose we use the inhomogeneous 
dynamical mean field theory (DMFT)\cite{Georges96,Bulla00,Potthoff99,Schwieger03,Okamoto04Nat,Okamoto04,Helmes08,Zenia09,Helmes08opt,Koga08,Koga09,Gorelik10} 
with the iterated perturbation
theory (IPT) as an impurity solver.\cite{Kajueter96,Saso99}
To be concrete, we analyze a Hamiltonian which has a topological 
non-trivial kinetic term as proposed for the HgTe/CdTe quantum wells
and include an additional local Hubbard interaction.
We show that the finite size gap is renormalized by interactions,
which can be naturally understood based on a Fermi-liquid picture.

In the second part of this article,
we discuss the paramagnetic Mott transition within the inhomogeneous
DMFT using the numerical renormalization group
(NRG)\cite{wilson1975,Bulla08} as an impurity solver.
We show that correlation effects are significantly stronger near the edge
sites, because the coordination number is smaller at the edges
than in the bulk, as has been discussed in correlated systems without
bulk
gaps.\cite{Bulla00,Potthoff99,Schwieger03,Okamoto04Nat,Okamoto04,Helmes08,Zenia09,Fujimoto04} 
Therefore, especially near the Mott transition,
the localization of the edge states around the edge sites
which is a fundamental property in TIs is weakened.
However, possible inter-edge Umklapp scattering due to finite size effects
does not seem to be important, and the "edge Mott insulating states"
are not found.
We show that even with strong correlations the ground state is a
topological insulating state possessing edge states,
which is adiabatically connected from the non-interacting system
for all interaction strengths smaller than the critical value.
When the interaction is stronger than the critical value,
a nearly homogeneous Mott insulating state is stabilized.
This is due to quantum tunneling of the electrons
from the bulk to the edges, which was previously discussed 
in correlated metals.
Finally, in the appendix of the article we present details on the NRG
calculations.

\section{Finite size effects}
\label{sec:finite}
\begin{figure}[tb]%
\includegraphics[width=0.8\hsize]{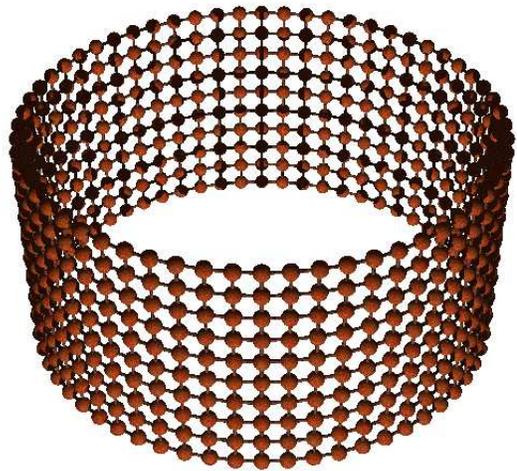}
\caption{(Color online) 
Ribbon geometry: square lattice with open boundaries in $y$-direction.}
\label{fig:ribbon}
\end{figure}

In this section, we discuss finite size effects,\cite{Zhou08,Linder09,Lu10,Wada11,Liu10,Potter10,Zhou11,YZhang10,Sakamoto10}
in two dimensional TIs of different sizes in the ribbon
geometry, see Fig. \ref{fig:ribbon}.
When its width $L_y$
is large compared to the characteristic
localization length of the edge states, $L_y\gg \xi_{\rm TI}$,
the edge states, which are localized on the two sides, are well separated.
Therefore, we can focus only on them and
study correlation effects by using the standard bosonization technique.
In this approach, forward scattering which has infrared
singurarities is fully taken into account, while other interaction
terms are perturbatively treated.
We note that this consideration is based on
the bulk-edge 
correspondence, which allows us to forget the details of the
bulk of the system.

For small system size where $L_y\gg \xi_{\rm TI}$ is not satisfied, 
the tunneling between the two
sides is no longer a perturbation and
the edge states are massive Dirac fermions.
In this case, we take into account electron correlations by means
of the inhomogeneous DMFT for a Hubbard model which will be introduced later.

We first discuss relatively large size TIs in the next section,
which is 
followed by the discussion of small systems in section \ref{sec:small}.

\subsection{Large systems: bosonization approach}
\label{sec:large}
It is known that in the presence of interactions,
each of the edge states, which are localized at the two sides, can be
described in terms of the helical Tomonaga-Luttinger liquid
(HTLL).\cite{Wu06,Xu06} 
However, when the two sides are correlated, the edge states 
are usual Tomonaga-Luttinger liquids (TLL) with spin-charge 
separation.\cite{Tanaka09,Hou09,Strom09,Teo09}
In this section, we investigate large size TIs by means of
the standard bosonization
approach,\cite{book:Giamarchi04,book:Tsvelik95} 
focusing on inter-edge tunneling processes.

Fermion operators for the edge states, localized at the 
edges $j=1,2$, are expressed as
\begin{eqnarray*}
\Psi_{jr\sigma}(x)=e^{irk_Fx}\psi_{jr\sigma}(x)
\end{eqnarray*}
for left($r=-$ or $L$) and right($r=+$ or $R$) 
movers with pseudo spins $\sigma
=\uparrow,\downarrow$, and $k_F$ being the Fermi wavenumber of the 
edge states.
The effective Hamiltonian for the two edge states is given by
\begin{align*}
H&=H_1+H_2+H_{ch}+H_{f}+H_{f^{\prime}}+H_{f^{\prime\prime}}
+H^{\prime},\\
H_1&=-iv\int dx [\psi_{1R\uparrow}^{\dagger}\partial_x\psi_{1R\uparrow}
-\psi_{1L\downarrow}^{\dagger}\partial_x\psi_{1L\downarrow}],\\
H_2&=+iv\int dx [\psi_{2L\uparrow}^{\dagger}\partial_x\psi_{2L\uparrow}
-\psi_{2R\downarrow}^{\dagger}\partial_x\psi_{2R\downarrow}],\\
H_{ch}&=g_{ch}\int dx[\rho_{1R\uparrow}\rho_{1R\uparrow}+
\rho_{1L\downarrow}\rho_{1L\downarrow}+(1\rightarrow 2)],\\
H_f&=g_f\int dx[\rho_{1R\uparrow}\rho_{1L\downarrow}+
\rho_{2L\uparrow}\rho_{2R\downarrow}],\\
H_{f^{\prime}}&=g_{f^{\prime}}\int dx[\rho_{1R\uparrow}\rho_{2L\uparrow}+
\rho_{1L\downarrow}\rho_{2R\downarrow}],\\
H_{f^{\prime\prime}}&=g_{f^{\prime\prime}}
\int dx[\rho_{1R\uparrow}\rho_{2R\downarrow}+
\rho_{1L\downarrow}\rho_{2L\uparrow}],
\end{align*}
where $\rho_{jr\sigma}=\psi_{jr\sigma}^{\dagger}\psi_{jr\sigma}$
and $v$ is the Fermi velocity for the edge states.
$H_{1,2}$ are kinetic terms for the edge states localized 
on edge-1 and edge-2 in the ribbon geometry, 
$H_{ch}\sim H_{f^{\prime\prime}}$
are forward scattering terms, and $H^{\prime}$ includes other
scattering processes.
In the terminology of usual spin $1/2$ fermions on a chain,
$H_{ch}, H_{f}, H_{f^{\prime}}$, and $H_{f^{\prime\prime}}$
correspond to $g_{4\parallel}, g_{2\perp}, g_{2\parallel}$,
and $g_{4\perp}$ terms, respectively.
We note that, in the present study, SU(2) symmetry for
spins is not required, and generally $g_{f}\geq g_{f^{\prime}},
g_{f^{\prime\prime}}$ holds.
Although the indices $R/L$ are redundant, 
we keep them so that the relation between the usual spin $1/2$ fermions
and the present system is clear.
We also note that the parameters in the above Hamiltonian
should be regarded as those including correlation effects in the
bulk system from which the effective one dimensional model is
deduced.

Possible scattering processes in $H^{\prime}$ are
\begin{widetext}
\begin{align*}
H^{\prime}&=H_{\lambda}+H_{\lambda^{\prime}}
+H_{hf}+H_{sf}+H_{ef}+H_u+H_{u^{\prime}},\\
H_{\lambda}&=
\lambda\int dx [\psi_{1R\uparrow}^{\dagger}\psi_{2L\uparrow}e^{-i2k_Fx}+
\psi_{2R\downarrow}^{\dagger}\psi_{1L\downarrow}e^{-i2k_Fx}
+({\rm h.c.})],\\
H_{\lambda^{\prime}}&=\lambda^{\prime}\int dx 
[\psi_{1R\uparrow}^{\dagger}\psi_{2R\downarrow}-
\psi_{1L\downarrow}^{\dagger}\psi_{2L\uparrow}
+({\rm h.c.})],\\
H_{hf}&=g_{hf}\int dx [\psi_{1R\uparrow}^{\dagger}
\psi_{2R\downarrow}^{\dagger}\psi_{1L\downarrow}\psi_{2L\uparrow}e^{-i4k_Fx}
+\psi_{2L\uparrow}^{\dagger}\psi_{1L\downarrow}^{\dagger}
\psi_{2R\downarrow}\psi_{1R\uparrow}e^{+i4k_Fx}],\\
H_{ef}&=g_{ef}\int dx[\psi_{2L\uparrow}^{\dagger}\psi_{2R\downarrow}^{\dagger}
\psi_{1L\downarrow}\psi_{1R\uparrow}
+\psi_{1R\uparrow}^{\dagger}\psi_{1L\downarrow}^{\dagger}
\psi_{2R\downarrow}\psi_{2L\uparrow}],\\
H_{sf}&=g_{sf}\int dx[\psi_{2L\uparrow}^{\dagger}\psi_{1R\uparrow}^{\dagger}
\psi_{1L\downarrow}\psi_{2R\downarrow}
+\psi_{2R\downarrow}^{\dagger}\psi_{1L\downarrow}^{\dagger}
\psi_{1R\uparrow}\psi_{2L\uparrow}],\\
H_u&=g_{u}\int dx [\psi_{1R\uparrow}^{\dagger}(x)
\psi_{1R\uparrow}^{\dagger}(x+a_0)
\psi_{1L\downarrow}(x+a_0)\psi_{1L\downarrow}(x)e^{-i4k_Fx}
+\psi_{2L\uparrow}^{\dagger}\psi_{2L\uparrow}^{\dagger}
\psi_{2R\downarrow}\psi_{2R\downarrow}e^{+i4k_Fx}+({\rm h.c.})],\\
H_{u^{\prime}}&=g_{u^{\prime}}\int dx [\psi_{1R\uparrow}^{\dagger}(x)
\psi_{1R\uparrow}^{\dagger}(x+a_0)
\psi_{2L\uparrow}(x+a_0)\psi_{2L\uparrow}(x)e^{-i4k_Fx}
+\psi_{2R\downarrow}^{\dagger}\psi_{2R\downarrow}^{\dagger}
\psi_{1L\downarrow}\psi_{1L\downarrow}e^{-i4k_Fx}+({\rm h.c.})],
\end{align*}
\end{widetext}
where $a_0$ is the lattice constant.
$H_{\lambda}$ is a spin-conserving tunneling term, and $H_{\lambda^{\prime}}$ 
is a spin-flip tunneling term which can be nonzero if the bulk Hamiltonian
does not have the spin rotation symmetry.
$H_{hf}, H_{ef}$, and $H_{sf}$ describe "helicity-flip",
"edge-flip" and "spin-flip" processes, respectively.
$H_u$ and $H_{u^{\prime}}$ are Umklapp terms.
$H_{hf}, H_{ef}$ and $H_{u^{\prime}}$ correspond to 
$g_{3\perp}, g_{1\perp}$ and $g_{3\parallel}$ terms in the spin $1/2$
fermions.
Obviously, they are invariant under the time-reversal transformation
$\psi_{jR/L\uparrow}(x)\rightarrow \psi_{jL/R\downarrow}(x)$ and
$\psi_{jR/L\downarrow}(x)\rightarrow -\psi_{jL/R\uparrow}(x)$.

To investigate the effects of $H^{\prime}$, we use the standard bosonization
approach.\cite{book:Giamarchi04,book:Tsvelik95}
Bosonization relations are
\begin{eqnarray*}
\psi_{jr\sigma}&=&\frac{U_{jr\sigma}}
{\sqrt{2\pi a_0}}e^{-ir\phi_{jr\sigma}},
\end{eqnarray*}
where $U_{jr\sigma}$ are the Klein factors.
We introduce
\begin{align*}
\phi_{c}&=(\phi_{1L\downarrow}+\phi_{1R\uparrow}
+\phi_{2L\uparrow}+\phi_{2R\downarrow})/2\sqrt{2},\\
\theta_{c}&=(\phi_{1L\downarrow}-\phi_{1R\uparrow}
+\phi_{2L\uparrow}-\phi_{2R\downarrow})/2\sqrt{2},\\
\phi_{s}&=(-\phi_{1L\downarrow}+\phi_{1R\uparrow}
+\phi_{2L\uparrow}-\phi_{2R\downarrow})/2\sqrt{2},\\
\theta_{s}&=(-\phi_{1L\downarrow}-\phi_{1R\uparrow}
+\phi_{2L\uparrow}+\phi_{2R\downarrow})/2\sqrt{2}\,.
\end{align*}
These variables satisfy 
\begin{equation*}
[\phi_{\mu}(x),\partial\theta_{\nu}(y)]=i\pi\delta_{\mu\nu}\delta(x-y),
\end{equation*}
and the canonical momentum is $\Pi_{\mu}=(1/\pi)\partial\theta_{\mu}$.

Then, the kinetic term in the Hamiltonian 
$H_K=H_1+H_2+H_{ch}+H_f+H_{f^{\prime}}+H_{f^{\prime\prime}}$
can be diagonalized and
the bosonized Hamiltonian is,
\begin{align*}
H_K&=H_{c}+H_{s},\\
H_{c}&=\frac{v_{c}}{2\pi}
\int dx\Bigl[\frac{1}{K_{c}}(\partial \phi_{c})^2
+K_{c}(\partial \theta_{c})^2\Bigr],\\
H_{s}&=\frac{v_{s}}{2\pi}
\int dx\Bigl[\frac{1}{K_{s}}(\partial \phi_{s})^2
+K_{s}(\partial \theta_{s})^2\Bigr],
\end{align*}
where
\begin{align*}
v_{c}&=[(v+\bar{g}_{ch}+\bar{g}_{f^{\prime\prime}}
+\bar{g}_{f}+\bar{g}_{f^{\prime}})\notag \\
&\quad\times(v+\bar{g}_{ch}+\bar{g}_{f^{\prime\prime}}
-\bar{g}_{f}-\bar{g}_{f^{\prime}})]^{1/2},\\
K_{c}&=\biggl[\frac{v+\bar{g}_{ch}+\bar{g}_{f^{\prime\prime}}
-\bar{g}_{f}-\bar{g}_{f^{\prime}}}{v+\bar{g}_{ch}+\bar{g}_{f^{\prime\prime}}
+\bar{g}_{f}+\bar{g}_{f^{\prime}}}\biggr]^{1/2},\\
v_{s}&=[(v+\bar{g}_{ch}-\bar{g}_{f^{\prime\prime}}
+\bar{g}_{f}-\bar{g}_{f^{\prime}})\notag\\
&\quad\times (v+\bar{g}_{ch}-\bar{g}_{f^{\prime\prime}}
-\bar{g}_{f}+\bar{g}_{f^{\prime}})]^{1/2},\\
K_{s}&=\biggl[\frac{v+\bar{g}_{ch}-\bar{g}_{f^{\prime\prime}}
+\bar{g}_{f}-\bar{g}_{f^{\prime}}}{v+\bar{g}_{ch}-\bar{g}_{f^{\prime\prime}}
-\bar{g}_{f}+\bar{g}_{f^{\prime}}}\biggr]^{1/2},
\end{align*}
with $\bar{g}_{ch}=g_{ch}/\pi$ and $\bar{g}=g/2\pi$ for the other terms.
$H_{c}$ and $H_{s}$ describe collective charge and 
spin excitations, respectively. 
As mentioned above, because $g_{f}\geq g_{f^{\prime}}$,
$K_{s}$ is not fixed to unity and generally $K_{s}\geq 1$.
If all the inter-edge correlated terms are neglected,
$H_{c}$ and $H_{s}$ are essentially identical, and 
the edge states are nothing but two copies of the HTLL localized 
at each edge.
We note that, if the bulk system is simply renormalized by
a renormalization factor $z$
in the presence of interactions,
$v$ is proportional to $z$ and $\bar{g}$ is to $z^2$.
Therefore, deviation of $K_{s,c}$ from unity would be suppressed by a
factor $z$.

$H^{\prime}$ is also bosonized as
\begin{widetext}
\begin{align*}
H_{\lambda}&=\frac{i\lambda}{\pi a_0}U_{1R\uparrow}U_{2L\uparrow}
\int dx \sin (\sqrt{2}(\phi_{c}+\phi_{s})-2k_Fx)
+\frac{i\lambda}{\pi a_0}U_{2R\downarrow}U_{1L\downarrow}
\int dx \sin (\sqrt{2}(\phi_{c}-\phi_{s})-2k_Fx),\\
H_{\lambda^{\prime}}&=
\frac{i\lambda^{\prime}}{\pi a_0}U_{2L\uparrow}U_{1L\downarrow}
\int dx \sin \sqrt{2}(\phi_{s}+\theta_{s})
+\frac{i\lambda^{\prime}}{\pi a_0}U_{1R\uparrow}U_{2R\downarrow}
\int dx \sin \sqrt{2}(\phi_{s}-\theta_{s}),\\
H_{hf}&=-u\frac{2g_{hf}}{(2\pi a_0)^2}
\int dx \cos (\sqrt{8}\phi_{c}-4k_Fx),\\
H_{ef}&=u\frac{2g_{ef}}{(2\pi a_0)^2}
\int dx \cos \sqrt{8}\phi_{s},\\
H_{sf}&=-u\frac{2g_{sf}}{(2\pi a_0)^2}
\int dx \cos \sqrt{8}\theta_{s},\\
H_{u}&=\frac{2g_{u}}{(2\pi a_0)^2}
\int dx [\cos (\sqrt{8}(\phi_{c}+\theta_{s})-4k_Fx)
+\cos (\sqrt{8}(\phi_{c}-\theta_{s})-4k_Fx)],\\
H_{u^{\prime}}&=\frac{2g_{u^{\prime}}}{(2\pi a_0)^2}
\int dx [\cos (\sqrt{8}(\phi_{c}+\phi_{s})-4k_Fx)
+\cos (\sqrt{8}(\phi_{c}-\phi_{s})-4k_Fx)),
\end{align*}
\end{widetext}
where $u=U_{2R\downarrow}U_{1R\uparrow}U_{1L\downarrow}U_{2L\uparrow}$.
$H_{\lambda}$ and Umklapp terms mix the charge and spin sectors.

The tunneling Hamiltonian $H_{\lambda}$ includes $2k_F$-oscillating factors,
because it describes hybridizations between edge states on the two sides
with the same wavenumber $k$.
In the following, we focus on a special filling, $k_F=0$, to
evaluate the finite-size gap induced by the tunneling Hamiltonian.
When $k_F=0$, the chemical potential is exactly at the Dirac point
of $H_{1,2}$ and
all the oscillating factors in the Hamiltonian become unity
$e^{i2k_Fx}=e^{i4k_Fx}=1$.
Then, we can evaluate the perturbations by looking at
their scaling dimensions $D_{\mu}$ for $H_{\mu}$,
\begin{align*}
D_{\lambda}&=(K_{c}+K_{s})/2,\\
D_{\lambda^{\prime}}&=(K_{s}+1/K_{s})/2,\\
D_{hf}&=2K_{c},\\
D_{ef}&=2K_{s},\\
D_{sf}&=2/K_{s},\\
D_{u}&=2(K_{c}+1/K_{s}),\\
D_{u^{\prime}}&=2(K_{c}+K_{s}).
\end{align*}
From these equations,
one sees that $D_{\lambda}$ is the smallest 
scaling dimension around the non-interacting point $K_{c}=K_{s}=1$,
and therefore, the spin-conserving tunneling $H_{\lambda}$ would be
the most relevant perturbation in $H^{\prime}$.
Therefore, the edge states become gapped by $H^{\prime}$, and 
they are no longer the TLL
in finite size systems when $k_F\sim 0$.
We note that, since the tunneling terms $H_{\lambda}$ and
$H_{\lambda^{\prime}}$ have non-zero conformal spins, they contribute to
renormalization group flows of $H_{hf}, H_{ef}$ and $H_{sf}$,
resulting in complicated flows.\cite{book:Giamarchi04,book:Tsvelik95,Brazovskii85,Yakovenko92,Kusmartsev92}
However,
because $H_{\lambda}$ is the most relevant perturbation,
we focus only on $H_{\lambda}$ 
and, in the following, simply neglect all the other perturbation
terms in $H^{\prime}$.

Under this approximation,
we can estimate the finite size gap by using the standard analysis.\cite{book:Giamarchi04}
When $\lambda$ is negative and large enough, 
conditions $\sqrt{2}(\phi_{c}\pm\phi_{s})\simeq \pi/2$ give the minimal energy
and the Hamiltonian would become
\begin{align*}
H_{\lambda} &\simeq \frac{\lambda}{\pi  a_0}2(\phi_{c}+\phi_{s})^2
+\frac{\lambda}{\pi  a_0}2(\phi_{c}-\phi_{s})^2\notag\\
&=\frac{2\tilde{\lambda}}{(\pi  a_0)^2}(
\phi_{c}^2+\phi_{s}^2),
\end{align*}
where $\tilde{\lambda}=2\pi  a_0 \lambda$.
Here, we have used a representation of the Klein factors where 
$U_{1R\uparrow}U_{2L\uparrow}=U_{2R\downarrow}U_{1L\downarrow}=-i$
can be satisfied, like
$U_{1R\uparrow}=\sigma_1\otimes\sigma_1,
U_{1L\downarrow}=1\otimes\sigma_2,
U_{2R\downarrow}=\sigma_3\otimes\sigma_1,
U_{2L\uparrow}=\sigma_2\otimes\sigma_1$, where $\{\sigma_i\}$ are
the Pauli matrices.
$\phi_{c}$ has been shifted by $\pi/2\sqrt{2}$.
Then, the spin and charge sectors are decoupled and the gaps
induced by $H_{\lambda}$ are respectively evaluated
as
\begin{align*}
\Delta_{s,c}^{\lambda}\sim w
\left( \frac{K_{s,c}\tilde{\lambda}}{w}\right)^{1/(2-D_{\lambda})},
\end{align*}
where $w$ is an effective band width of the edge states which
would correspond to the bulk gap $\Delta_{\rm TI}$.
$\Delta_{s,c}$ are enhanced by the forward scattering. 
We stress that, even for the weak interactions $K_{s,c}\simeq 1$, 
both of the spin and charge sectors are gapped by the tunneling $H_{\lambda}$.
Therefore, the TLL description fails for the low energy 
edge states in the finite size TIs when $k_F\simeq 0$.
Nevertheless, correlation effects peculiar to one-dimensional systems 
appear as a difference between the charge gap and the spin gap, 
$\Delta_c^{\lambda} \neq \Delta_s^{\lambda}$, which characterizes
the spin-charge separation. 
We stress that the charge gap can be much smaller than the spin gap 
in strongly correlated regions provided that the edge states can be treated as one-dimensional systems.
It should be an interesting issue to experimentally detect 
the spin-charge separation raised by the one-dimensional correlation 
effect from the measurement  of the Hall effect and the spin Hall effect 
which are governed by the edge states.

In the above derivation of $\Delta_{s,c}$, we have simply neglected
other interaction terms in $H^{\prime}$.
However, they can be important in a strongly correlated regime
where $K_{s,c}$ is away from unity.
As mentioned above, $H_{\lambda^{\prime}}, H_{sf}, H_{u}$ 
describe spin flip processes which are difficult 
to take into accout in numerical calculations.
On the other hand, the inter-egdge interactions 
$H_{hf}, H_{u^{\prime}}$ are spin-conserving Umklapp terms
and might lead to the possible "edge Mott insulating states".
The effects of the inter-edge Unklapp scattering are discussed in
Sec. \ref{sec:DMFT+NRG}.

At the end,
we note that all the parameters in the expressions for the gaps, $K_{s,c}, 
\tilde{\lambda}$ and $w$,
are derived from the two dimensional bulk system
and would be renormalized by the interactions in the bulk.
This renormalization cannot be calculated from the one dimensional model,
and rather should be included in the initial values of the parameters.

In the next section, we analyze small TIs by an alternative approach
which is based on a Fermi-liquid picture for the TIs.

\subsection{Small systems: DMFT+IPT calculations}
\label{sec:small}
When the system size is small enough so that the tunneling between the
edges is no longer a small perturbation,
analyses based on well-defined massless modes are not applicable anymore.
In this section,
we consider correlation effects in small size systems
whose edge states are massive Dirac fermions.

To be concrete, we consider the following Hamiltonian
defined on a square lattice in the ribbon geometry
(Fig.~\ref{fig:ribbon}),
\begin{equation}
H=H_{\rm BHZ}+H_{\rm int}\,,
\label{eq:BHZ+U}
\end{equation}
where $H_{\rm BHZ}$ is topologically equivalent to the 
Hamiltonian for the HgTe/CdTe quantum wells.\cite{BHZ06,Konig07,Hasan10,Qi11}
$H_{\rm int}$ is an interaction term between the electrons.
The Hamiltonian contains two orbitals described by
$C_i=(c_{i1\uparrow},c_{i2\uparrow},c_{i1\downarrow},c_{i2\downarrow})^T$
where the indices $1, 2 (\uparrow, \downarrow)$ label orbitals (pseudo spins).
The kinetic energy can be written as
\begin{widetext}
\begin{subequations}
\begin{align}
H_{\rm BHZ}&=
\sum_{ij}C^{\dagger}_{i}\hat{H}_{ij}C_{j},
\label{eq:BHZ}\\
\hat{H}_{ij}&=\left[
\begin{array}{cc}
{\mathcal H}_{ij} & 0 \\
0 & {\mathcal H}^{\ast}_{ij}
\end{array}
\right],\\
{\mathcal H}_{ij}&=\left[
\begin{array}{cc}
M_0\delta_{ij}-t(\delta_{i,j\pm \hat{x}}+ \delta_{i,j\pm \hat{y}}) 
& t^{\prime} [i(\delta_{i,j+\hat{x}}-\delta_{i,j-\hat{x}})
+\delta_{i,j+\hat{y}}-\delta_{i,j-\hat{y}}] \\
t^{\prime} [i(\delta_{i,j+\hat{x}}-\delta_{i,j-\hat{x}})
+\delta_{i,j-\hat{y}}-\delta_{i,j+\hat{y}}]
& -M_0\delta_{ij}+t(\delta_{i,j\pm \hat{x}}+ \delta_{i,j\pm \hat{y}})
\end{array}
\right].
\end{align}
\end{subequations}
\end{widetext}
The periodic boundary condition is imposed for the $x$-direction,
while the system has a finite width $L_y$ for the $y$-direction.
As will be discussed later, this Hamiltonian is particle-hole symmetric
when the filling is $n_1+n_2=1$ per spin-direction.
Regarding the interaction term,
we consider the simplest Hamiltonian
\begin{align}
H_{\rm int}=U\sum_{il}n_{il\uparrow}n_{il\downarrow},
\end{align}
where $n_{il\sigma}=c_{il\sigma}^{\dagger}c_{il\sigma}$.
The total Hamiltonian is a Hubbard model with inter-site 
inter-orbital hopping $t^{\prime}$ corresponding to the spin-orbit
(SO) interaction in the 
present article.
In the following, we take $t$ as the energy unit $t=1$, and
the SO interaction $t^{\prime}$ is fixed
to be $t^{\prime}=0.25$. Furthermore, we set the filling to
$n_{i1}+n_{i2}=1$ per spin-direction for all sites.
We also fix $L_y=10$ and $M_0=-1.0$, if not specified, for which
the finite size gap is $\Delta_{\rm f}\sim 0.1\Delta_{\rm TI}$
in the absence of the interaction.
As discussed in the previous section, in finite size TIs,
infrared singularities are cut off and descriptions based on
the TLL fail to capture gapped edge states when their Fermi
wavenumber is small.
Instead, a Fermi-liquid picture of correlated band insulators
would give a correct understanding where 
the edge states are massive Dirac fermions, as will be discussed
in the following sections.

\subsubsection{Non-interacting case}
\label{sec:Non-int}
\begin{figure}[tb]%
\includegraphics[width=\linewidth]{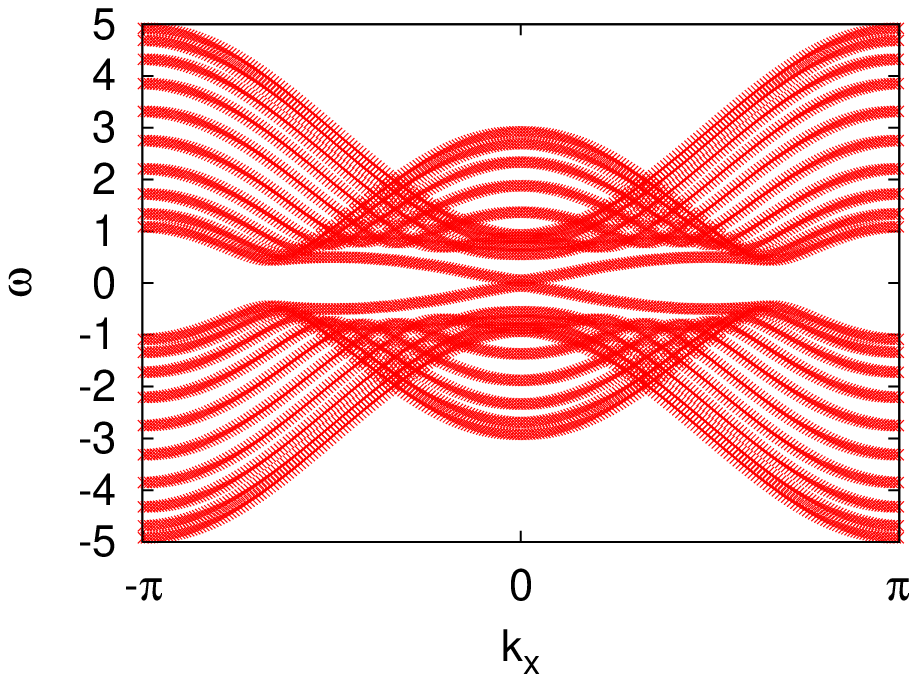}\\
\includegraphics[width=\linewidth]{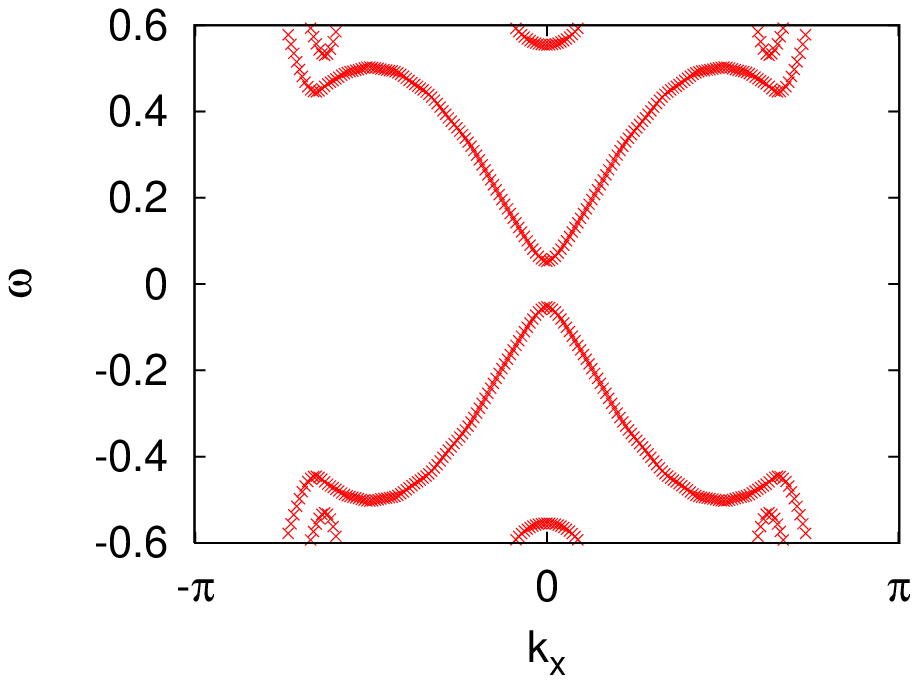}
\caption{(Color online) The dispersion relation for 
$M_0=-1.0$. The edge states spectrum is enlarged in the 
lower panel
so that the finite size gap can be seen.}
\label{fig:disp}
\end{figure}

\begin{figure}[tb]%
\includegraphics[width=\linewidth]{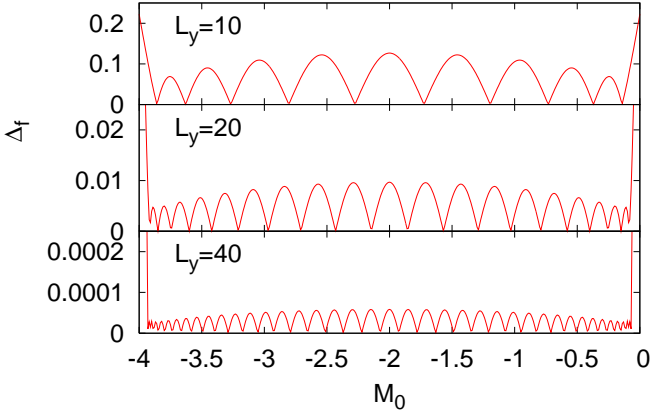}
\caption{(Color online) The finite size gap as a function of $M_0$
for $L_y=10, 20, 40$ from the top to the bottom.
For $0<M_0<4$, $\Delta_{\rm f}$ depends on $M_0$ in the same
way as for $-4<M_0<0$.}
\label{fig:gap0}
\end{figure}

Before discussing correlation effects, 
we want to show some basic properties of the non-interacting system.
In Fig. \ref{fig:disp}, 
the dispersion relation for $M_0=-1.0$ 
calculated from $H_{\rm BHZ}$ is shown.
The dispersion is particle-hole symmetric and
edge states exist for 
$0<|M_0|<4$. The bulk gap is $\Delta_{\rm TI}\simeq 1.0$
corresponding to $t^{\prime}=0.25$.
The edge states have a small but finite gap as discussed in 
previous studies,\cite{Zhou08, Linder09, Lu10, Wada11,Liu10,Potter10,Zhou11,
YZhang10,Sakamoto10}
The finite-size gap $\Delta_{\rm f}$ depends on the parameter $M_0$ and it is
oscillating when $M_0$ is tuned as shown in Fig. \ref{fig:gap0}
for $L_y=10, 20$ and 40.
The number of the oscillation for $0<|M_0|<4$
is $2(L_y-1)$, which is related to
the number of the eigenstates of $H_{\rm BHZ}$ for a fixed $k_x$.
This oscillation is specific to lattice systems,
and we have to be careful about this when discussing correlation effects
in the following sections where oscillations in $\Delta_{\rm f}$
can also be seen.
Oscillations in bulk gaps with respect to system sizes have been
discussed in many articles and oscillatory dimensional crossovers are seen
in some classes of TIs and TSCs.\cite{Zhou08, Linder09, Lu10,
  Wada11,Liu10,Potter10,Zhou11,YZhang10,Sakamoto10} 
However, in our study, the non-interacting system remains topologically
non-trivial as long as $0< |M_0|< 4$.
$\Delta_{\rm f}$ is exponentially decreasing as $L_y$ is increased, 
and its size dependence is roughly
\begin{align*}
\Delta_{\rm f}&\simeq \Delta_{\rm f0}\exp 
\left(-L_y/\xi_{\rm TI}^{(0)} \right),\\
\xi_{\rm TI}^{(0)} &\sim t^{\prime}/M_0,
\end{align*}
where $\xi_{\rm TI}^{(0)}$ characterizes the localization of the edge states
around the edge sites in the non-interacting system.
$\Delta_{\rm f0}$ is defined by a combination of the 
parameters in $H_{\rm BHZ}$. 
Exact expressions for these quantities in the continuum limit
can be found in \textcite{Zhou08}.

Naive expectation based on the Fermi-liquid picture together with
an assumption that
 in the presence of the interaction $U$ the
low energy states are renormalized by a single renormalization
factor $z$, results in a finite-size gap
\begin{subequations}
\begin{align}
\Delta_{\rm f0}(U)&\simeq z\Delta_{\rm f0}(U=0),
\label{eq:expect1}\\
\xi_{\rm TI}(U) &\sim zt^{\prime}/zM_0\sim \xi_{\rm TI}^{(0)}
\label{eq:expect2}.
\end{align}
\end{subequations}
Renormalizations of $t^{\prime}$ and $M_0$ would cancel out and
$\xi_{\rm TI}(U)$ would be equal to $\xi_{\rm TI}^{(0)}$.
Therefore, our simplest expectation is that 
the amplitude of the finite size gap is renormalized by $z$ while
$\xi_{\rm TI}$ is not, although the legitimacy of the above assumption
is not so clear.
In the following sections, we investigate correlation effects
and show that the naive expectation actually holds in the
presence of the Hubbard interaction within the DMFT.

\subsubsection{Inhomogeneous DMFT+IPT}

In this section, we discuss the correlation effects by means
of the inhomogeneous DMFT+IPT.
The inhomogeneous DMFT is an extension of the DMFT \cite{Georges96} 
to inhomogeneous
systems and has been applied to, e.g.  metallic
surfaces,\cite{Potthoff99,Schwieger03} 
heterostructures,\cite{Okamoto04Nat,Okamoto04,Helmes08,Zenia09}
and optical lattices.\cite{Helmes08opt,Koga08,Koga09,Gorelik10}
We note that, in the present study, the edge states for small $L_y$
are massive and not the massless HTLLs as discussed in the previous
sections. 
The whole two dimensional system including the
edge degrees of freedom is supposed to be
a gapped Fermi-liquid.
Therefore, we expect that inhomogeneous DMFT can capture the main properties of 
the edge states in the presence of interactions, 
although it is known that DMFT fails to describe the low energy 
TLL behavior in one dimension.\cite{Capone04}

To calculate the selfenergy, we solve the following
self consistent equations for $y=1, 2, \cdots, L_y$,
\begin{eqnarray*}
{\mathcal G}^{-1}_{aa^{\prime}}(\omega,y)
=\left[\frac{1}{L_x}\sum_{k_x}G
(\omega,k_x,y,y)\right]^{-1}_{aa^{\prime}}
+\Sigma_{aa^{\prime}}(\omega,y),
\end{eqnarray*}
where $a=(l\sigma)$, ${\mathcal G}$ is the cavity Green's function and
$L_x$ is the number of $k_x$ points, which is taken to be
sufficiently large.
The lattice Green's function $G_{aa^{\prime}}(\omega,k_x,y,y^{\prime})$
is a $4L_y\times 4L_y$ matrix, where the system size is fixed as
$L_y=10$ in the present study.
We note that $G$ is spin-diagonal and
$\sum_{k_x}G_{l\sigma,l^{\prime}\sigma}(\omega,k_x,y,y)=
\sum_{k_x}G_{l\bar{\sigma},l^{\prime}\bar{\sigma}}(\omega,k_x,y,y)$ 
is satisfied in our ribbon geometry, 
and therefore we often suppress the spin indices throughout this
article.

The selfenergy of the effective impurity problem is 
calculated by means of the IPT which can compute the selfenergy as a function
of real frequency $\omega$.\cite{Kajueter96,Saso99}
Within the IPT, the selfenergy is evaluated in the following way:
The first order contribution is
$\Sigma^{(1)}_{l\sigma,l^{\prime}\sigma^{\prime}}(y)
=Un_{l\sigma}(y)\delta_{ll^{\prime}}\delta_{\sigma\sigma^{\prime}}$ where 
$n_{l\sigma}(y)$ is the electron density for orbital $l$ and
spin $\sigma$ at site $y$.
We note that $\sum_{l\sigma}n_{l\sigma}(y)=1$ for all the sites.
The second order is
\begin{eqnarray*}
\Sigma^{(2)}_{d_1d_2}(\omega)&=&-\frac{1}{2\pi^3}
\Gamma^{a_1c_1}_{d_2b_2}\Gamma^{d_1b_1}_{a_2c_2}\nonumber\\
&\times&\!\!\!
\int\!\! d\varepsilon d\varepsilon^{\prime} d\varepsilon^{\prime\prime}
{\rm Im}\tilde{{\mathcal G}}_{a_2a_1}(\varepsilon)
{\rm Im}\tilde{{\mathcal G}}_{b_2b_1}(\varepsilon^{\prime})
{\rm Im}\tilde{{\mathcal G}}_{c_2c_1}(\varepsilon^{\prime\prime})\nonumber\\
&\times&\frac{1}{\omega-\varepsilon+\varepsilon^{\prime}
-\varepsilon^{\prime\prime}}
\Bigl
\{f(\varepsilon)[1-f(\varepsilon^{\prime})]f(\varepsilon^{\prime\prime})
\nonumber\\
&&+[1-f(\varepsilon)]f(\varepsilon^{\prime})[1-f(\varepsilon^{\prime\prime})]\Bigr\}\,,
\end{eqnarray*}
where $a,b,c,d=(l\sigma)$ and $\Gamma$ is an antisymmetric
bare vertex. $f$ is the Fermi distribution function.
In the present model,
$\Gamma_{l\sigma,l\bar{\sigma}}^{l\sigma,l\bar{\sigma}}=U,
\Gamma_{l\sigma,l\bar{\sigma}}^{l\bar{\sigma},l\sigma}=-U$,
and all the other elements are zero.
The Green's function $\tilde{\mathcal G}$ is given by
$\tilde{\mathcal G}^{-1}_{l\sigma l^{\prime}\sigma^{\prime}}
={\mathcal G}^{-1}_{l\sigma l^{\prime}\sigma^{\prime}}
-Un_{l\sigma}\delta_{l l^{\prime}}\delta_{\sigma\sigma^{\prime}}$.
The selfenergy is interpolated from the weak coupling to the atomic 
and high energy limit in the IPT calculations,  
\begin{eqnarray*}
\Sigma_{l\sigma,l^{\prime}\sigma}(\omega)
&=&\Sigma^{(1)}_{l\sigma,l^{\prime}\sigma}
+\frac{A_{ll^{\prime}}\Sigma^{(2)}_{l\sigma,l^{\prime}\sigma}(\omega)}
{1-B_{ll^{\prime}}\Sigma^{(2)}_{l\sigma,l^{\prime}\sigma}(\omega)},
\end{eqnarray*}
where $A_{ll^{\prime}}=\delta_{ll^{\prime}}n_l(1-n_l/2)/
\tilde{n}_l(1-\tilde{n}_l/2)+(1-\delta_{ll^{\prime}})$,
$B_{ll^{\prime}}=\delta_{ll^{\prime}}2(1-n_l)/
\tilde{n}_l(1-\tilde{n}_l/2)$, and
$\tilde{n}_l$ is calculated from $\tilde{\mathcal G}$.
We use a simplified scheme which has been successfully 
applied to correlated metals away from the half filling,
in which the effective chemical potential is simply determined from
$\Sigma(\omega=0)=\Sigma^{(1)}(0)$.\cite{Saso99}

\begin{figure}[tb]%
\includegraphics[width=\linewidth]{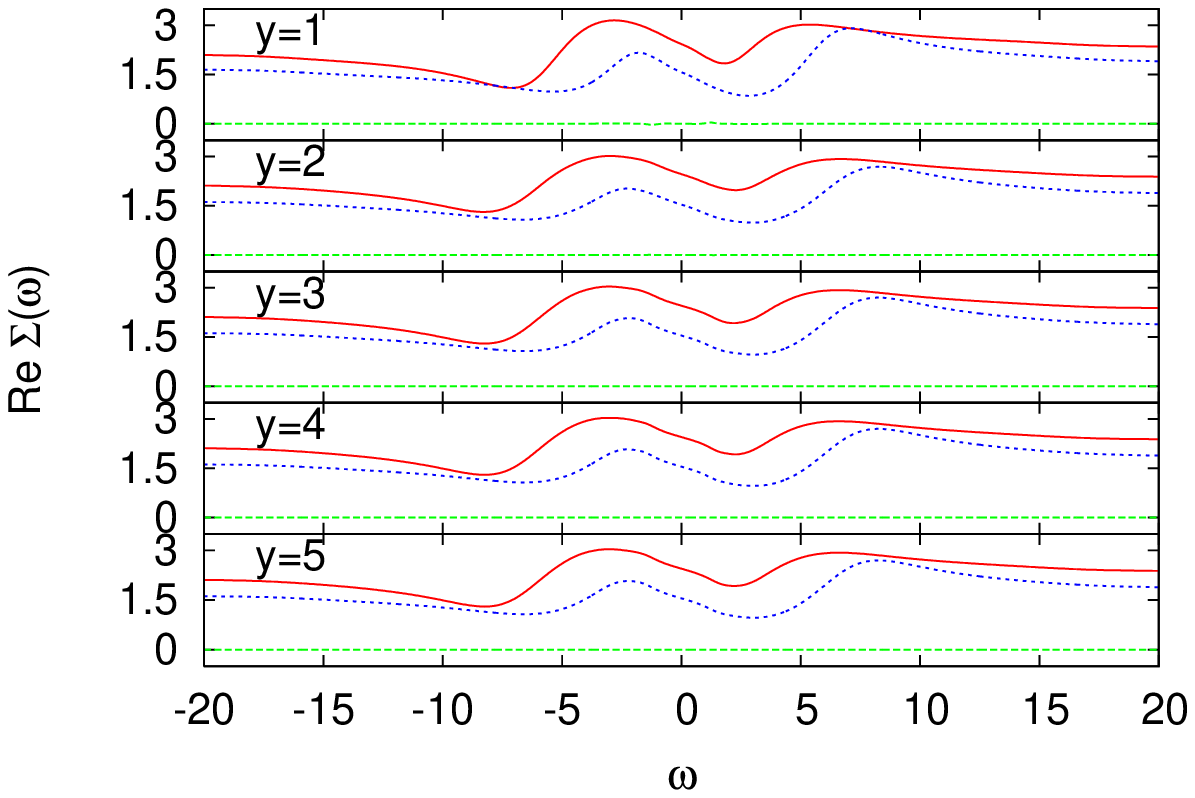}\\
\includegraphics[width=\linewidth]{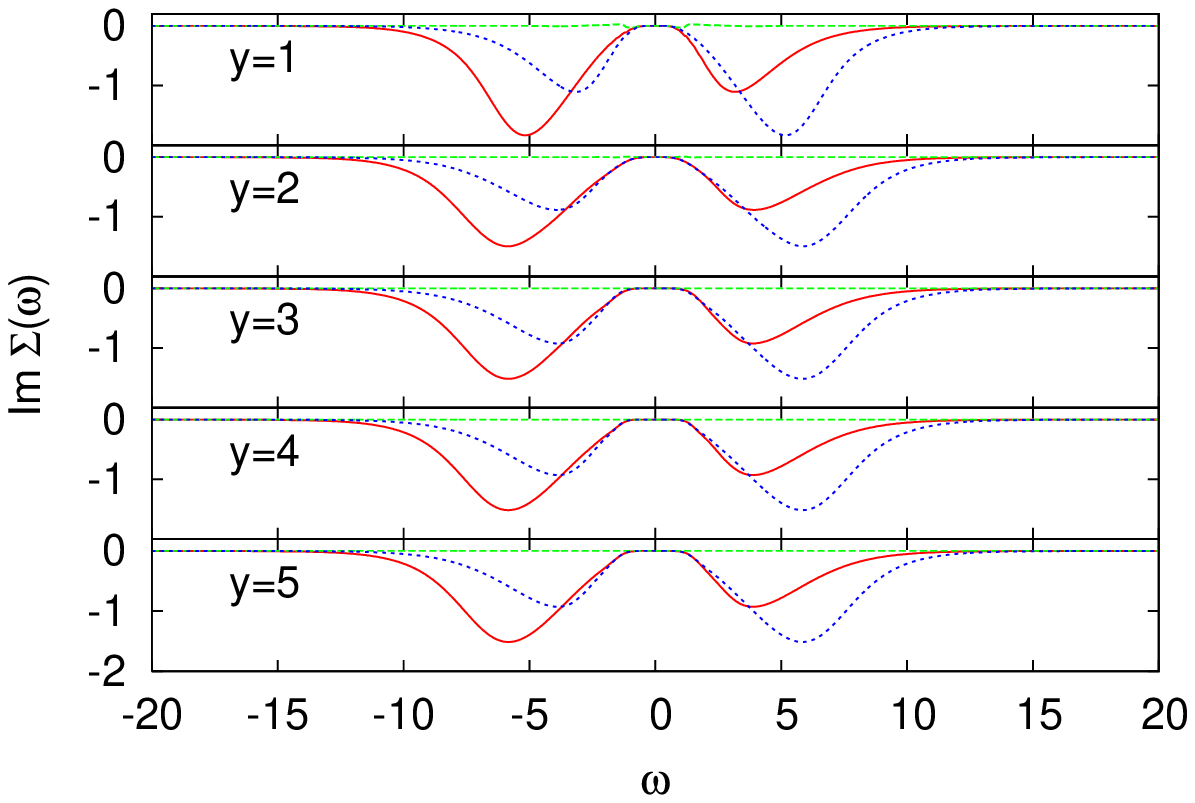}
\caption{(Color online) Selfenergy for $U=4$ and $L_y=10$.
Red, green, blue curves are $\Sigma_{11}, \Sigma_{12,21}$ and
$\Sigma_{22}$, respectively. 
}
\label{fig:SigU4_IPT}
\end{figure}

\begin{figure}[tb]%
\includegraphics[width=\linewidth]{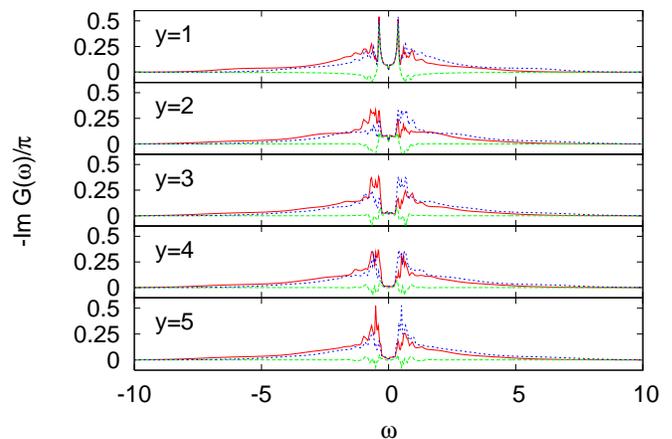}
\caption{(Color online) The local Green's functions for $y=1\sim5$
from the top to the bottom when $U=4$ and $L_y=10$.
The red, green, and blue curves are for $G_{11}, G_{12,21}$,
and $G_{22}$, respectively.
}
\label{fig:LDOSU4}
\end{figure}

\begin{figure}[tb]%
\includegraphics[width=\linewidth]{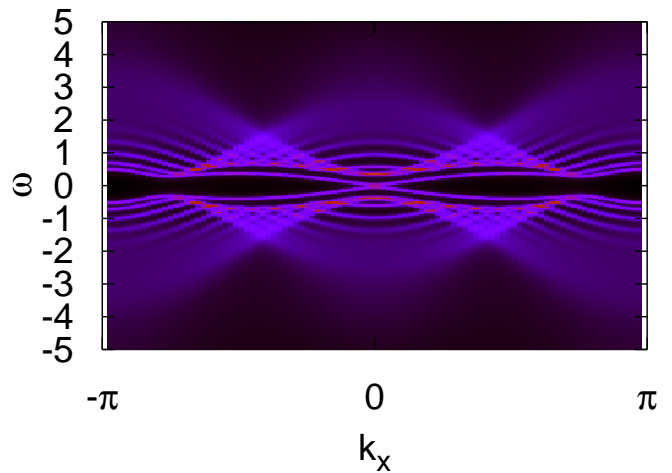}
\caption{(Color online) The energy spectrum $A(k_x,\omega)$ for $U=4$ and
  $L_y=10$.
}
\label{fig:kxU4_IPT}
\end{figure}

In Figs. \ref{fig:SigU4_IPT} and \ref{fig:LDOSU4}, 
we show the selfenergy and the local Green's functions for half of the
system width for $U=4$.
As will be discussed later, the local selfenergies for the other half
are 
${\rm Im}\Sigma_{11,22}(\omega,L_y-y+1)={\rm Im}\Sigma_{11,22}(\omega,y)$ and
${\rm Im}\Sigma_{12,21}(\omega,L_y-y+1)=
-{\rm Im}\Sigma_{12,21}(\omega,y)$.
The local Green's functions show the same symmetry properties.
We see that, the diagonal elements of the
selfenergy
are similar to those of simple band insulators;
e.g. ${\rm Im}\Sigma_{11,22}(\omega)$ seem to have gap like structures.
They behave like 
${\rm Im}\Sigma_{11,22}(\omega,y)\simeq -b(y) \omega^2$ for small
$\omega$, where 
the prefactors $b(y)$ are much smaller than those in metallic systems.
This is because the density of states has 
only small weight around $\omega=0$ which corresponds to the edge states.
On the other hand,
although ${\rm Im}G_{12,21}$ is 
comparable to ${\rm Im}G_{11,22}$ in
magnitude at least for $y=1$ around $\omega\sim\Delta_{\rm TI}$, 
$\Sigma_{12,21}$ is much smaller than $\Sigma_{11,22}$.
In Fig. \ref{fig:kxU4_IPT},
we also show the energy spectrum for $U=4$ defined by
$A(k_x,\omega)=-(1/L_y)\sum_{y}{\rm Im}
[G_{11}(k_x,y,\omega)+G_{22}(k_x,y,\omega)]$.
Compared to Fig. \ref{fig:disp}, the edge states around $\omega=0$
are renormalized and high energy structures are smeared.
We note that, although the massive Dirac states are topologically protected,
namely they are free from the single particle backscattering and
the intra-edge Umklapp scattering in the spin conserved systems,
they have a long but finite lifetime due to the forward scattering.
A direct comparison with a non-perturbative impurity solver, the NRG, 
is shown in Appendix \ref{ch:Disc}.

\begin{figure}[tb]%
\includegraphics[width=\linewidth]{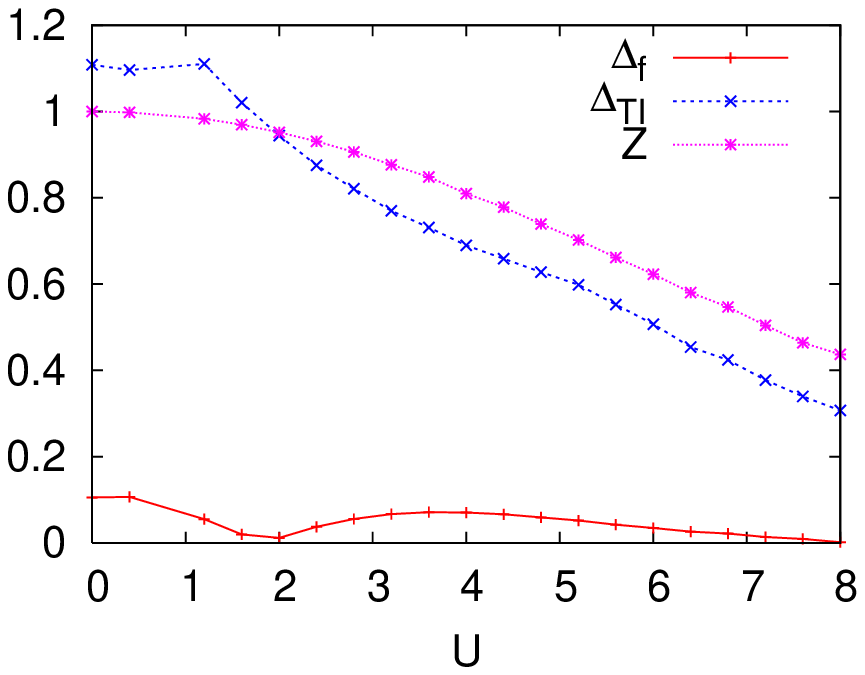}\\
\includegraphics[width=\linewidth]{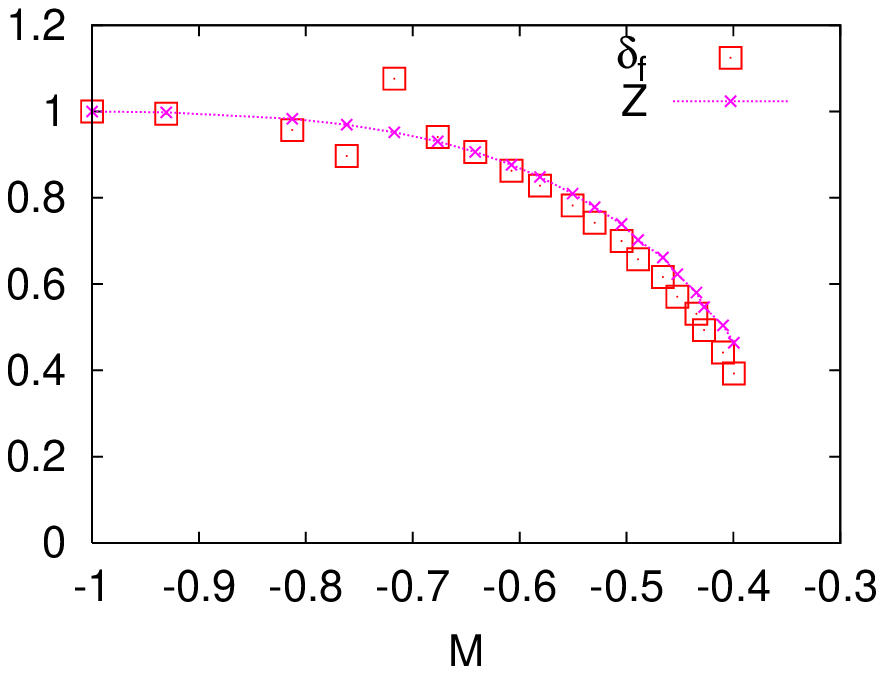}
\caption{(Color online) (Upper panel) The finite size gap $\Delta_{\rm f}$, 
the bulk gap $\Delta_{\rm TI}$, and the $y$-averaged
renormalization factor ${\mathcal Z}$ as functions of $U$
for $L_y=10$.
(Lower panel) The ratio of the finite size gap $\delta_{\rm f}$ 
and ${\mathcal Z}$ as functions of ${\mathcal M}$. 
}
\label{fig:gap_L10}
\end{figure}

The finite size gap $\Delta_{\rm f}$ 
as a function of $U$ can be calculated from
an effective Hamiltonian
\begin{eqnarray*}
\bar{H}(\omega)&=&\hat{H}+\Sigma(\omega).
\end{eqnarray*}
For $|\omega|<\Delta_{\rm TI}$, ${\rm Im}\Sigma(\omega)$ is
much smaller than that for $t^{\prime}=0$ and can be neglected
when evaluating the finite size gap.
We can identify the positions of zeros in $\omega-\bar{H}(\omega)$,
which correspond to the energy eigenvalues of the edge states.
As shown in the upper panel of Fig. \ref{fig:gap_L10}, 
$\Delta_{\rm f}$ is an oscillating function of $U$.
This oscillation has its origin in the non-interacting
system in Fig. \ref{fig:gap0},
because constant terms in ${\rm Re}\Sigma_{11,22}$ play the same role as $M_0$
in the inverse of the Green's functions.
The bulk gap $\Delta_{\rm TI}$ is simply estimated from the
next lowest excitation energy of $\bar{H}$ at $k_x=0$,
and ${\mathcal Z}
=(1/L_y)\sum_y z(y)$ is also shown in the figure.
We note that, compared to $\Delta_{\rm TI}(U)$,
the renormalization of $\Delta_{\rm f}$ is rather moderate,
because $z(y)$, which is defined by the derivative of $\Sigma$ at $\omega=0$,
is larger than that at $\omega\simeq \Delta_{\rm TI}$.
To extract the correlation effects from this oscillating $\Delta_{\rm f}$,
we replot it as a function of a parameter ${\mathcal M}$ 
corresponding to $M_0$.
We define
\begin{eqnarray*}
{\mathcal M}(U)&=&{\mathcal M}_0
+\frac{1}{2L_y}\sum_y\Bigl[{\rm Re}\Sigma_{11}(\omega=0,y)\nonumber\\
&&-{\rm Re}\Sigma_{22}(\omega=0,y)\Bigr],\\
\delta_{\rm f}({\mathcal M})&=&
\Delta_{\rm f}({\mathcal M}(U))/\Delta_{\rm f}(M_0;U=0)|_{M_0={\mathcal M}},
\end{eqnarray*}
where ${\mathcal M}_0=-1.0$.
In the definition of $\delta_{\rm f}$, we have identified
the non-interacting parameter $M_0$ as ${\mathcal M}$.
In the lower panel in Fig. \ref{fig:gap_L10}, 
we show $\delta_{\rm f}({\mathcal M})$ and
${\mathcal Z}({\mathcal M})$.
We see that $\delta_{\rm f}({\mathcal M})$ and 
${\mathcal Z}(\mathcal M)$ well coincide,
which means that our naive expectation Eqs. (\ref{eq:expect1}) and
(\ref{eq:expect2}) based on the Fermi-liquid picture
actually holds within the DMFT+IPT calculations.
We note that, for weak interactions, site dependence in
$\Sigma(y)$ is not strong and $\Delta_{\rm f}$ can be
well understood in terms of their averages, ${\mathcal M}$ and
${\mathcal Z}$.
If the site dependence in $\Sigma$ is significant,
this interpretation is no longer applicable.

\begin{figure}[tb]%
\includegraphics[width=\linewidth]{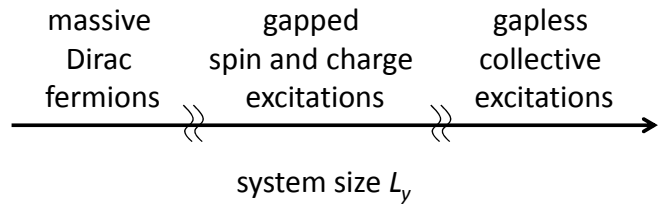}
\caption{The elementary excitations in two dimensional TIs for different 
system sizes when the Coulomb interactions are weak.}
\label{fig:excitations}
\end{figure}

In this section, we have discussed the correlation effects 
for small size systems, while the previous section was
dedicated to rather large systems.
For these two limiting situations, the physical pictures
are very different; the edge states are renormalized massive Dirac fermions
for small systems, instead of
the gapped spin and charge collective modes for large systems.
However, it would be possible to smoothly connect them to each
other as the  
system size $L_y$ is tuned.
If one starts from a small size where the single particle spectrum has
a gap,
one should calculate the susceptibilities to see the gaps 
in the spin and charge collective excitations. 
These gaps in the susceptibilities should be evaluated with the
vertex corrections being taken into account.
Although such calculations are beyond the present DMFT calculations,
a crossover would take place at some length scale of the system.
Such crossovers between two or three dimensional physics and 
one dimensional physics with respect to energy scales are also 
seen in ladder systems. \cite{book:Giamarchi04}
In Fig. \ref{fig:excitations}, we summarize the elementary excitations in two
dimensional TIs for different system sizes when the Coulomb interactions are
weak.
In the limit $L_y\rightarrow\infty$ where the inter-edge
correlation is negligible, the "gapless
collective excitations" correspond to the HTLL.

\section{Mott transition}
\label{sec:DMFT+NRG}
In this section, we discuss the paramagnetic Mott transition
for a topologically nontrivial system
by using the inhomogeneous DMFT.
We use the same model Hamiltonian as in the previous section,
Eq. (\ref{eq:BHZ+U}), and suppress possible magnetic instabilities.
In previous
studies\cite{Pesin10,Gurarie11,Yamaji11,Fidkowski10_1,Fidkowski10_2,Turner11}
it was discussed
that for TIs and TSCs with strong interactions novel ground states
might be realized, in which the bulk 
remains gapped with a non-trivial topological number
while the edge states are suppressed by interaction effects.
Such states are called ``edge Mott insulating states'',
and can be explicitly defined in terms of 
the Green's functions.\cite{Gurarie11}
Let us briefly consider a semi-infinite TI, which is built up as 
a heterostructure of the vacuum and the TI having one boundary.
In the absence of interactions, poles are supposed to be
in the  
Green's functions corresponding to the edge states, 
while in the presence of interactions there might be additional zeros
as was discussed by \textcite{Gurarie11}.
In the latter case, the edge states are suppressed and the
the ground state can be characterized as an 
``edge Mott insulating state''.
Although this discussion cannot be directly
applied to finite size systems with two boundaries, 
the selfenergy must have poles inside the bulk gap to realize
the edge Mott insulating state, which characterizes the TIs with
Mott insulating edge states.
In this scenario, the finite size gap would be replaced
by the Mott gap as interactions are increased.
We note that,
to obtain the edge Mott insulating states, the Umklapp scattering for the
edge states is essential.
Although, the Umklapp scattering of the
edges at the same side is absent in our model, there are possible
Umklapp scattering processes 
which involve both sides.
In the edge state Hamiltonians introduced in 
Sec. \ref{sec:large} these processes are given by
$H_{hf}$ and $H_{u^{\prime}}$, while $H_{u}$
is absent.
Such scattering processes might drive the edge states into a Mott
insulating state, 
keeping the bulk a topological insulator.
We note that, while $H_{u^{\prime}}$ is characterized by scattering at
different sites, 
$H_{hf}$ describes scattering at a single site.
Because there is some overlap of the edge state wave functions
which are localized mainly at opposite sides, 
effects of $H_{hf}$ are taken into account within the inhomogeneous
DMFT.
In addition, when $\Delta_{\rm f}/\Delta_{\rm TI}$ is not so small and
$U$ is large compared to $\Delta_{\rm TI}$, 
the inter-edge Umklapp scattering is expected to be important.

\subsection{Symmetries in Green's function and selfenergy}
\label{sec:GandS}

Before discussing results obtained with inhomogeneous DMFT, 
we discuss some identity relations in $G$ and $\Sigma$ 
arising from the
particle-hole symmetry in the Hamiltonian Eq. (\ref{eq:BHZ+U}).
These identities are quite useful for checking numerical results
and simplifying calculations, thus improving the accuracy.

The particle-hole transformation, defined by a unitary operator 
${\mathcal C}$, is given as
\begin{align*}
c_{i1\sigma}&\rightarrow c^{\dagger}_{i2\sigma},\\
c_{i2\sigma}&\rightarrow c^{\dagger}_{i1\sigma}.
\end{align*}
The condition that 
the Hamiltonian should be invariant under this transformation
determines the chemical potential.
The interaction in the model is given by $H_{\rm
  int}=U\sum_{il}n_{il\uparrow}n_{il\downarrow}$. Thus, the chemical
potential should be $\mu=U/2$ so that
particle-hole symmetry is conserved.
We note that our model belongs to the class DIII-Hamiltonians
which describe the $p\pm ip$ superconductors,\cite{Schnyder08,
  Schnyder09, Kitaev09} such as the Rashba 
superconductors, \cite{SF09}
if we identify our particle-hole symmetry ${\mathcal C}$
as emergent particle-hole symmetry in superconductors.\cite{Turner11}
This correspondence would allow us to interpret the off-diagonal selfenergy
$\Sigma_{12,21}$ as an anomalous selfenergy of the superconductivity
if the emergent particle-hole symmetry is kept in the presence of
interactions.

From the particle hole symmetry we can directly derive identities in
$G(\omega)$ and $\Sigma(\omega)$. 
The local retarded Green's function
for site $i$ 
satisfies
\begin{align}
\left[
\begin{array}{cc}
G^R_{11}(\omega,i)&G^R_{12}(\omega,i)\\
G^R_{21}(\omega,i)&G^R_{22}(\omega,i)
\end{array}\right]
=
\left[
\begin{array}{cc}
-G^R_{22}(-\omega,i)^{\ast}&-G^R_{12}(-\omega,i)^{\ast}\\
-G^R_{21}(-\omega,i)^{\ast}&-G^R_{11}(-\omega,i)^{\ast}
\end{array}\right].
\label{eq:sym_G}
\end{align}
We note that the inverse of the matrix 
$G(\omega)=[G_{ll^{\prime}}(\omega,i,j)]$
also satisfies the corresponding relations, since 
$G(\omega)^{-1}\cdot G(\omega)=G(\omega)\cdot G(\omega)^{-1}=
{\bf 1}$.

Thus, the selfenergy for a fixed filling fulfills similar relations, which
are directly derived from 
\begin{align*}
\Sigma^R(\mu;\omega,i)&=[G_{0}^R(\mu;\omega)]^{-1}_{ii}
-[G^R(\mu;\omega)]^{-1}_{ii}\\
&=\left[[G_{0}^R(\mu;\omega)]^{-1}_{ii}
-[G^R(\mu_0;\omega)]^{-1}_{ii}\right]+(\mu-\mu_0),
\end{align*}
where each of the non-interacting and interacting 
local Green's functions, the site-diagonal elements of the matrices
$G_{0}^{R}(\mu_0;\omega)$ and 
$G^{R}(\mu;\omega)$,
satisfies Eq.(\ref{eq:sym_G}). The chemical potential is $\mu_0=0$
and $\mu=U/2$. 
Therefore, the local selfenergy should satisfy 
\begin{align}
&\left[
\begin{array}{cc}
\Sigma^{R}_{11}(-\omega,i)-\mu&\Sigma^{R}_{12}(-\omega,i)\\
\Sigma^{R}_{21}(-\omega,i)&\Sigma^{R}_{22}(-\omega,i)-\mu
\end{array}\right]\notag\\
&\qquad =
\left[
\begin{array}{cc}
-\Sigma^{R}_{22}(\omega,i)^{\ast}+\mu&-\Sigma^{R}_{12}(\omega,i)^{\ast}\\
-\Sigma^{R}_{21}(\omega,i)^{\ast}&-\Sigma^{R}_{11}(\omega,i)^{\ast}+\mu
\end{array}\right].
\label{eq:sym_S}
\end{align}
We see that the real part of $\Sigma_{12,21}(\omega)$ is 
an odd function of $\omega$ and the imaginary part is an even function,
and especially ${\rm Re}\Sigma_{12,21}(\omega=0)=0$.
Looking at the inverse of the Green's function 
$G^{-1}=[\omega-\hat{H}-\Sigma]$, 
the off-diagonal selfenergy $\Sigma_{12,21}(\omega,i)$ 
can be interpreted as a local hybridization between the orbitals,
which does not add to the inter-site inter-orbital hopping
$t^{\prime}$ in the Hamiltonian $H$.
In this interpretation, $\Sigma_{12,21}(\omega,y)$ are regarded as the
induced "$s$-wave" hybridization between 
the orbitals arising from the "$p$-wave" hybridization $t^{\prime}$ 
in the presence of the on-site interaction.\cite{remark}
The odd frequency dependence in ${\rm Re}\Sigma_{12,21}(\omega)$
is related to the odd parity character of the 
inter-orbital hopping 
$t^{\prime}$.

The odd parity $\Sigma_{12,21}$ in our system
is in contrast to those in trivial band insulators,
where inter-orbital hopping is
local, $s$-wave-like. 
It can be shown for this model that
off-diagonal elements behave like $\Sigma_{12}(\omega,i)
=\Sigma_{12}(-\omega,i)^{\ast}$.
Therefore, ${\rm Re}\Sigma_{12,21}(\omega=0,i)\neq0$ can
add to the local hybridization,  
which suggests that the bulk gap 
grows continuously as $U$ is
increased. This has been shown in the previous studies
for topologically trivial correlated band
insulators.\cite{Moeller99,Fuhrmann06,Kancharla07} 
In the context of superconductivity, this would correspond to
the BCS-BEC crossover for the $s$-wave pairing states.\cite{Bauer09,Bauer09PRB}

We also note that this argument can easily be generalized to
the non-local selfenergy, and the off-diagonal elements in
the present model satisfy $\Sigma_{12}(\omega,i,j)=-\Sigma_{12}
(-\omega,j,i)^{\ast}$ for sites $i,j$.
At $\omega=0$, $\Sigma_{12}(0,i,j)$ has the same symmetry as
the inter-site inter-orbital hopping and enhances it.
We again stress that the odd-parity symmetries in the off-diagonal selfenergy
are characteristic of our model.
However, constant terms in
${\rm Re}\Sigma_{12}(\omega,i,j)$ do not appear in the lowest order in $U$, 
and ${\rm Re}\Sigma_{12}(0,i,j)$
comes only from higher order terms in $U$.
Although ${\rm Re}\Sigma_{12}(0,i,j)$ 
is expected to be small in the present model,
sufficiently large ${\rm Re}\Sigma_{12}(0,i,j)$ might lead to
a crossover with continuous growth of the bulk gap 
in the presence of on-site interactions.
However, previous studies on the Kane-Mele Hubbard model
taking into account spatial correlations
show a renormalization of the bulk gap,\cite{Yu11}, which implies that
in our model
$\Sigma_{12}(0,i,j)$ for $i\neq j$ is not dominant, too.

Other relations in $G$ and $\Sigma$ 
corresponding to an inversion transformation
along the $y$-axis also exist.
The inversion transformation ${\mathcal I}$ is defined by
\begin{eqnarray*}
c_{il\sigma}&\rightarrow&c_{\bar{i}l\sigma},
\end{eqnarray*}
where $\bar{i}=(-x,L_y+1-y)$.
Under this transformation, the Hamiltonian satisfies
${\mathcal I}H(t^{11,22}_{ij},t^{12,21}_{ij})
{\mathcal I}^{-1}
=H(t^{11,22}_{ij},-t^{12,21}_{ij})$.
Because $G_{11,22}$ are even with respect to $t_{12,21}$ while
$G_{12,21}$ are odd,
the following relations hold:
\begin{align}
\left[
\begin{array}{cc}
G^R_{11}(\omega,i)&G^R_{12}(\omega,i)\\
G^R_{21}(\omega,i)&G^R_{22}(\omega,i)
\end{array}\right]
&=
\left[
\begin{array}{cc}
G^R_{11}(\omega,\bar{i})&-G^R_{12}(\omega,\bar{i})\\
-G^R_{21}(\omega,\bar{i})&G^R_{22}(\omega,\bar{i})
\end{array}\right],\\
\left[
\begin{array}{cc}
\Sigma^{R}_{11}(\omega,i)&\Sigma^{R}_{12}(\omega,i)\\
\Sigma^{R}_{21}(\omega,i)&\Sigma^{R}_{22}(\omega,i)
\end{array}\right]
&=
\left[
\begin{array}{cc}
\Sigma^{R}_{11}(\omega,\bar{i})
&-\Sigma^{R}_{12}(\omega,\bar{i})\\
-\Sigma^{R}_{21}(\omega,\bar{i})&
\Sigma^{R}_{22}(\omega,\bar{i})
\end{array}\right].
\end{align}
We note that, although $G_{12,21}(\omega,y)$ 
are comparable to $G_{11,22}(\omega,y)$
in magnitude around the edges, they are strongly suppressed in the bulk.
They become exactly zero at the center of the system if $L_y$ is odd,
as can be seen in the above identities.

\subsection{DMFT + NRG calculations}
\begin{figure}[tb]%
\includegraphics[width=\linewidth]{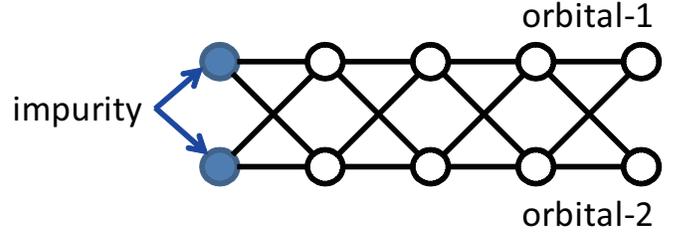}
\caption{(Color online) A schematic picture of the NRG chain Hamiltonian of
the effective impurity Anderson model.}
\label{fig:IAM}
\end{figure}

We will now discuss the correlation effects in small size
TIs from moderate to
strong coupling within the inhomogeneous DMFT using the numerical
renormalization group (NRG) \cite{wilson1975,Bulla08} as an impurity solver.
The system size is fixed as $L_y=10$. If nothing else is stated, we use a discretization parameter
$\Lambda=3.0$ and $N_K=5000$ states are kept in the calculations. 
The effective impurity Anderson model for a single lattice site reads
\begin{eqnarray*}
H_{\rm IAM}(y)&=&H_c(y)+H_{hyb}(y)+H_{imp}(y),
\label{eq:IAM}\\
H_c&=&\sum_{ms,ll^{\prime}\sigma}a^{\dagger}_{l\sigma}(ms)
g_{ll^{\prime}\sigma}(ms)a_{l^{\prime}\sigma}(ms),\\
H_{hyb}&=&\sum_{ms,ll^{\prime}\sigma}
\Bigl[f_{l\sigma}^{\dagger}h_{ll^{\prime}\sigma}(ms)
a_{l^{\prime}\sigma}(ms)\\
&&+a^{\dagger}_{l^{\prime}\sigma}(ms)h_{ll^{\prime}\sigma}^{\ast}(ms)
f_{l\sigma}\Bigr],\\
H_{imp}&=&\sum_{l\sigma}E^f_{l\sigma}
f_{l\sigma}^{\dagger}f_{l\sigma}
+H_{int}[f_{l\sigma}^{\dagger},f_{l\sigma}],
\end{eqnarray*}
where the parameters are determined self-consistently within the DMFT
loop and depend on the $y$-index of the lattice.
Details for the derivation of the parameters are given in Appendix
\ref{chp:imp} and \ref{app:chain}.
The Hamiltonian can be rewritten as a linear chain as shown in
Fig. \ref{fig:IAM}. 
To examine the effects induced by strong correlations,
the NRG is expected to be a very accurate impurity 
solver, especially near the Fermi energy.\cite{wilson1975,Bulla08}
However, due to the SO interaction and the resulting
inter-orbital hopping in the chain Hamiltonian, truncation effects
can become large in these calculations. 
Therefore, the NRG must
also be 
considered as an approximation even around the Fermi energy.
(A discussion on truncation effects is given in Appendix \ref{ch:Disc}.)
Furthermore, instead of using the
usual logarithmic broadening of the delta-peaks in the Green's
functions calculated by the NRG,\cite{Bulla08} we use a Lorentzian
broadening. The reason for this choice is, that the usual logarithmic
broadening seems to be more fragile towards breaking of the
symmetries discussed above. Using a Lorentzian broadening turned out
to be much more stable.
Nevertheless, NRG, being non-perturbative, 
can provide results for selfenergies and spectral functions, from
which the ground state properties 
can be estimated.
To our knowledge, the NRG has never been used for systems with inter-site
inter-orbital hopping.

\begin{figure}[tb]%
\includegraphics[width=\linewidth]{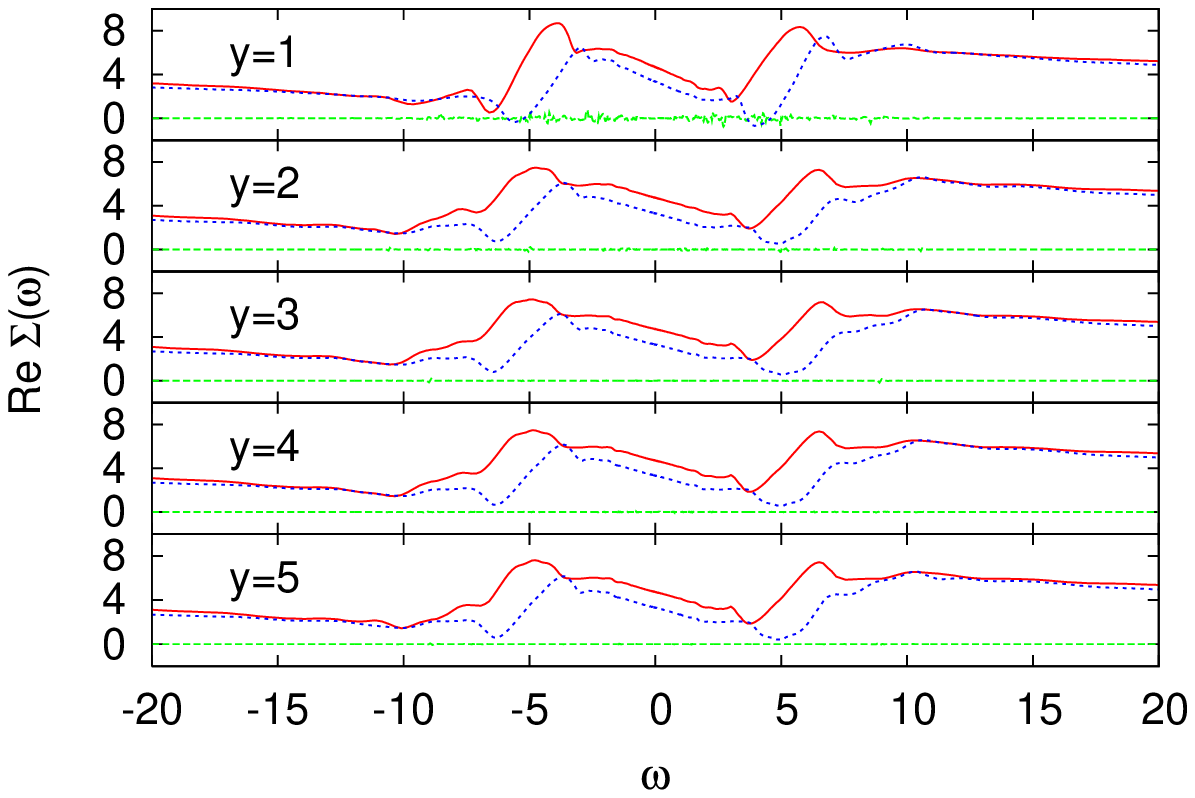}\\
\includegraphics[width=\linewidth]{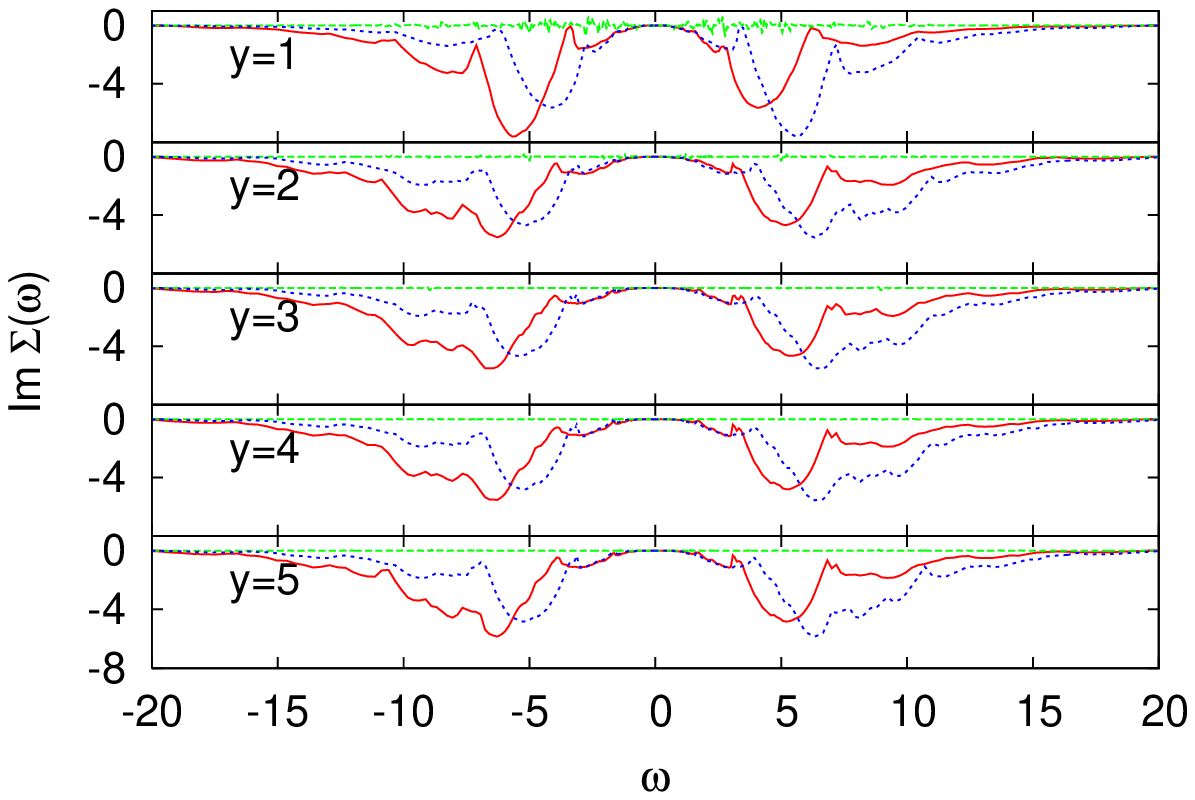}
\caption{(Color online) Selfenergy for $U=8$ and $L_y=10$.
Red, green, blue curves are $\Sigma_{11}, \Sigma_{12,21}$ and
$\Sigma_{22}$, respectively. 
}
\label{fig:SigU8}
\end{figure}

As discussed in the beginning of this section,
the structures of the selfenergy can distinguish
usual topological insulating states from possible
edge Mott insulating states.
Therefore, we first look at the $\omega$-dependence of the selfenergy.
In Fig. \ref{fig:SigU8}, $\Sigma(\omega,y)$ for $U=8$ is shown as an
example.
A general feature is that
the off-diagonal elements are much smaller than the diagonal  
elements.
Furthermore, the selfenergy for any site $y$ can be well
approximated by $\Sigma(\omega,y)
=[\Sigma(0,y)+(1-z(y)^{-1})\omega-ib(y)\omega^2]$,
where $z(y)=[1-\partial \Sigma(y)/\partial \omega
]_{\omega=0}^{-1}$. This behavior of $\Sigma_{11,22}$
is similar to those in correlated metals.
Therefore, the result can be characterized as a renormalized
gapped Fermi-liquid with renormalization factor $z(y)$ and
quasi-particle broadening $\sim b(y)\omega^2$.
There are no
characteristic features in $\Sigma(\omega)$,
such as isolated poles at $|\omega|<\Delta_{\rm TI}$,
which would drive the edge states into Mott insulating states.
Moreover, the prefactors $b(y)$ are significantly smaller than
those in metals, because the density of states around $\omega\simeq0$
is suppressed by the bulk gap.
This simple $\omega$-dependence of $\Sigma(\omega)$ agrees with
the previous DMFT+IPT calculations.
DMFT+NRG confirms that this dependence actually holds
for larger values of $U$ than the band width $W$.
The edge states always remain renormalized massive Dirac fermions
as long as $U$ is smaller than the critical value of the Mott transition
$U_c$.
The edge Mott insulating states are not found in our calculations.
Possible inter-edge Umklapp scattering does not seem to
have significant effects on the edge states, 
although its coupling constant $g_{hf}$
might be large compared to the gap width $\Delta_{\rm f, TI}$
when $U\gg \Delta_{\rm f, TI}$.
We note that all correlation effects can be characterized by a
single quantity $U/W$, and concerning the Mott insulator transition,
$U\sim \Delta_{\rm TI}$ does not introduce a new characteristic energy scale,
although $\Delta_{\rm TI}$ can seemingly characterize the
effective band width of the edge states.

Although we cannot find the edge Mott insulating
states in the present calculations, there is
another possibility for Mott insulating states to appear.
In inhomogeneous strongly correlated systems,
such as optical lattices with harmonic potentials,\cite{Helmes08opt}
Mott insulating behaviors in some domains of the systems have been found
at finite temperatures. 
For these states, the site dependence of local correlations is most
important.

\begin{figure}[tb]%
\includegraphics[width=\linewidth]{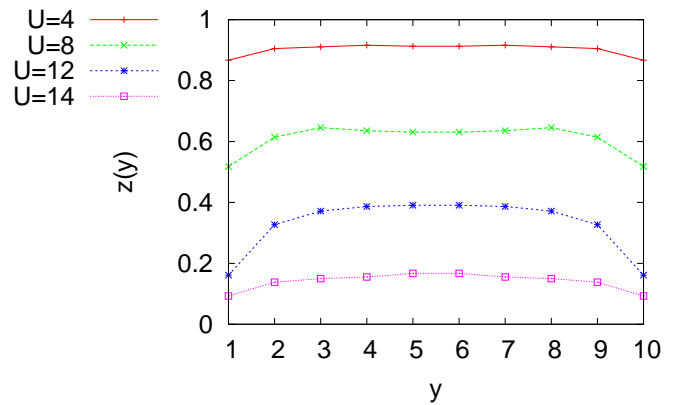}
\caption{(Color online) The renormalization factor $z(y)$ for several 
values of $U$ for $L_y=10$.
}
\label{fig:z}
\end{figure}

In the following,
we will discuss the spatial dependence of correlation effects
by looking at the renormalization factors $z(y)$.
In Fig. \ref{fig:z}, we compare $z(y)$ for several values of the
interaction strength $U$.
While for weak interactions, the $y$-dependence of $z(y)$ is moderate,
for strong interactions, e.g. $U=12$ and $U=14$, which 
are close to the critical value for the Mott transition,
$z(y)$ is largely suppressed, especially at the edge sites.
That can be understood by recalling that the coordination
number at the edge sites is   
smaller than in the bulk, which restricts the electron
motions at the edges.
Strong correlation effects at boundaries 
due to the reduction of the coordination number have been
discussed in the literature for various systems, such as
metallic surfaces,\cite{Potthoff99,Schwieger03}
heterostructures,\cite{Okamoto04Nat,Okamoto04,Helmes08,Zenia09} 
optical lattices,\cite{Helmes08opt,Koga08,Koga09,Gorelik10}
and spin systems.\cite{Fujimoto04}
Even though the electrons are largely renormalized especially around the edges,
$z(y)$ is nonzero for all the sites and the bulk gap due to the
SO interaction remains finite.
This state can be characterized as a renormalized TI, and therefore
can be adiabatically connected 
to the non-interacting TI without closing the gap. 
In this sense, the system remains topologically
unchanged for all interaction strengths $U<U_c$, where $U_c$ is the
critical interaction strength for the Mott transition.

\begin{figure}[tb]%
\includegraphics[width=\linewidth]{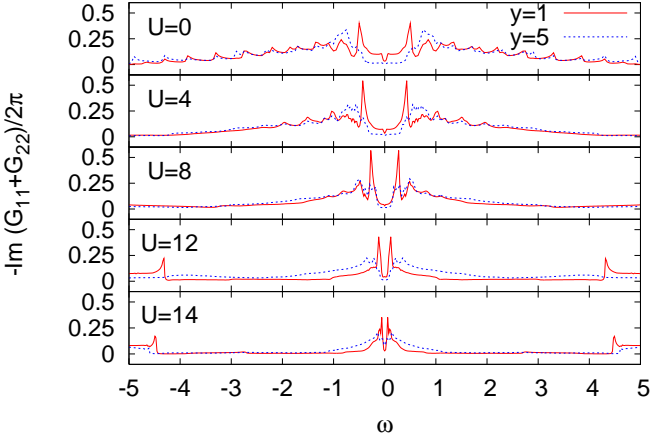}
\caption{(Color online) The local density of states at $y=1$ (red curve) and 
$y=5$ (blue curve) for several values of $U<U_c$.
}
\label{fig:ImG_U0U14}
\end{figure}

The orbital averaged local density of states $-(1/2\pi){\rm Im}(G_{11}
+G_{22})$ for small $\omega$ directly shows the renormalization effects
as seen in Fig. \ref{fig:ImG_U0U14}.
Small dips inside the bulk gap correspond to the finite size gap, and
finite values for $|\omega|<\Delta_{\rm TI}$
are due to the massive Dirac edge states.
As has been discussed in Sec. \ref{sec:small}, both $\Delta_{\rm TI}$
and $\Delta_{\rm f}$ are renormalized in a similar way.

For larger values of $U$, $U> 14$, 
the renormalization factors become zero $z(y)=0$ for all $y$,
signaling Mott insulating states for all sites.
Our results are qualitatively equivalent to the results for the single
band Hubbard model 
where both edges and bulk are metallic without interactions.
Although $z(y)$ decreases especially around the edge sites, 
as has also been found in the previous studies \cite{Bulla00,Potthoff99,Schwieger03,Okamoto04Nat,Okamoto04,Helmes08,Zenia09}, 
$z(y)$ vanishes for all $y$ at the same time at zero temperature, and 
a nearly homogeneous Mott insulating state is stabilized.
This can be explained by quantum tunneling of electrons
from the bulk to the edges. If the bulk stays metallic,
then, the edges are imposed to be metallic even though $z$ at the edges
is strongly suppressed.

In the TIs, on the other hand, the bulk is gapful, and therefore, one
could expect that 
the quantum tunneling of electrons is diminished, which 
possibly leads to
inhomogeneous solutions where only the edges are gapped with 
a Mott gap $\Delta_{\rm MI}$.
Such solutions
could be interpreted as heterostructures of 
Mott insulators and TIs.
Nevertheless, the Mott insulating sites near the edges 
can reduce the effective system size of the TI,
leading to an enhanced finite size gap.

However, as mentioned above, inhomogeneous Mott insulating
states are not obtained in our zero temperature calculations.
The absence of such inhomogeneous solutions can already be understood 
within the linearized DMFT.
As was shown in Figs. \ref{fig:SigU4_IPT} and \ref{fig:SigU8},
the local selfenergy can be approximated as $\Sigma_{11,22}(\omega,
y)=\Sigma_{11,22}(0,y)+(1-z(y)^{-1})\omega+{\mathcal O}(\omega^2)$ 
for small $\omega$, and 
$\Sigma_{12,21}$ is much smaller than $\Sigma_{11,22}$ in 
magnitude and can be neglected.
In this approximation, $\Sigma(\omega)$ has the same structure as a
simple two orbital 
metallic system, and the discussions in the previous
studies\cite{Bulla00,Potthoff99,Schwieger03,Helmes08,Zenia09} 
can be easily applied to the system.
Within the linearized DMFT, inhomogeneous solutions
such as $z(y=1,L_y)=0$ and $z(1<y<L_y)\neq 0$ are not stable
provided that $\Sigma_{12,21}$ is negligible.
Because the bulk gap is generated by the inter-site inter-orbital hopping,
the gap generating mechanism extends to the edge by
electron motions, making the whole system topological insulating.
In other words, the system is never cut by $z=0$ at specific sites
when the interaction strength is increased.
This argument is based upon the linearized DMFT equations and is a 
direct consequence of
the simple structure of the local selfenergy $\Sigma_{ll^{\prime}}
(\omega,y)\simeq[\Sigma_{ll^{\prime}}(0,y)+
(1-z(y)^{-1})\omega]\delta_{ll^{\prime}}$ which can be 
checked in the DMFT+NRG and DMFT+IPT calculations.
We also note that, even if spatial correlations are 
taken into account within cluster extensions of the DMFT, 
Mott transitions for large size systems 
are expected to be homogeneous.\cite{Ishida10}

However, especially at the edge sites, 
the Fermi-liquid behavior in our system
would be fragile against temperatures, and
Mott insulating domains could be seen at finite temperatures
as it was discussed in the literature.\cite{Zenia09,Helmes08opt}
One can expect that there are temperature ranges where the peak structures 
around $\omega\sim 0$ in
the local density of states at the edge sites disappear, while 
they survive in the bulk.
In such a case, Mott insulating domains could grow from
the edges to the bulk as $U$ is increased.
Finite temperature studies will be discussed in the future.

\begin{figure}[tb]%
\includegraphics[width=\linewidth]{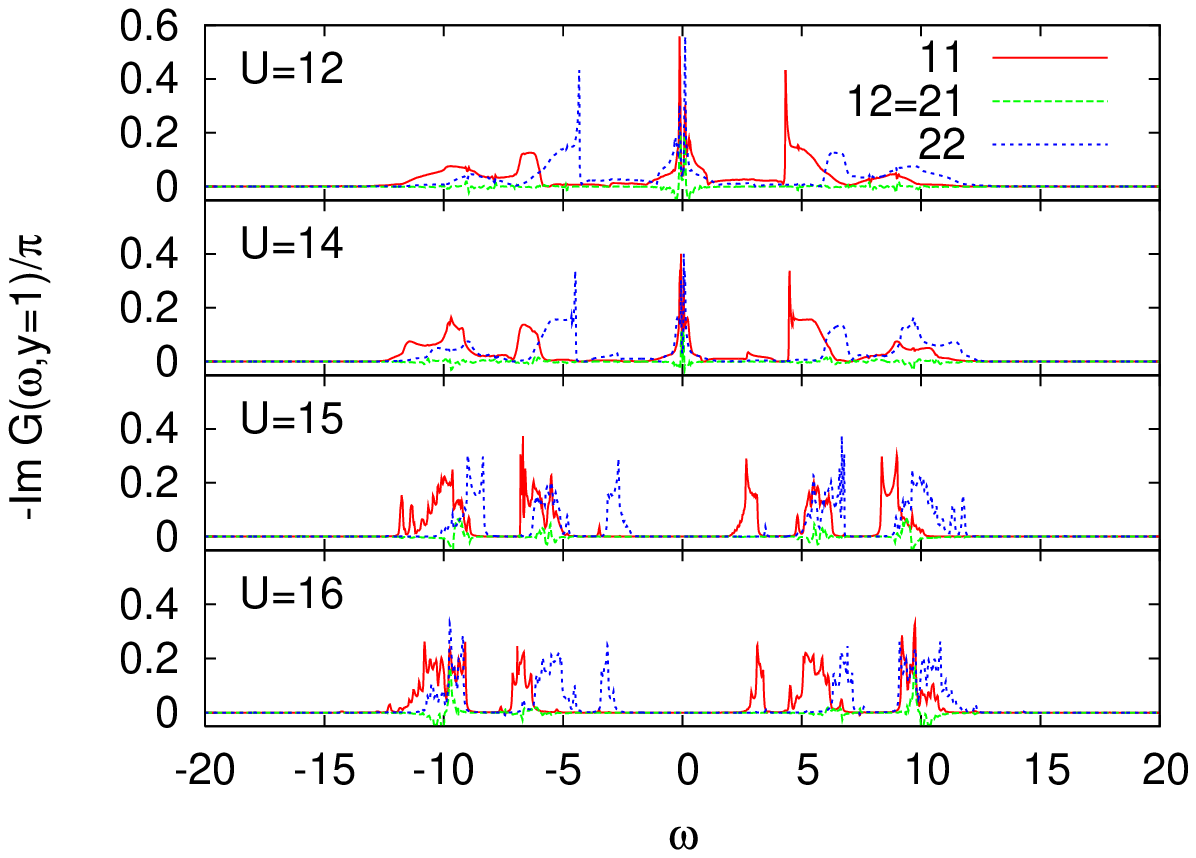}\\
\includegraphics[width=\linewidth]{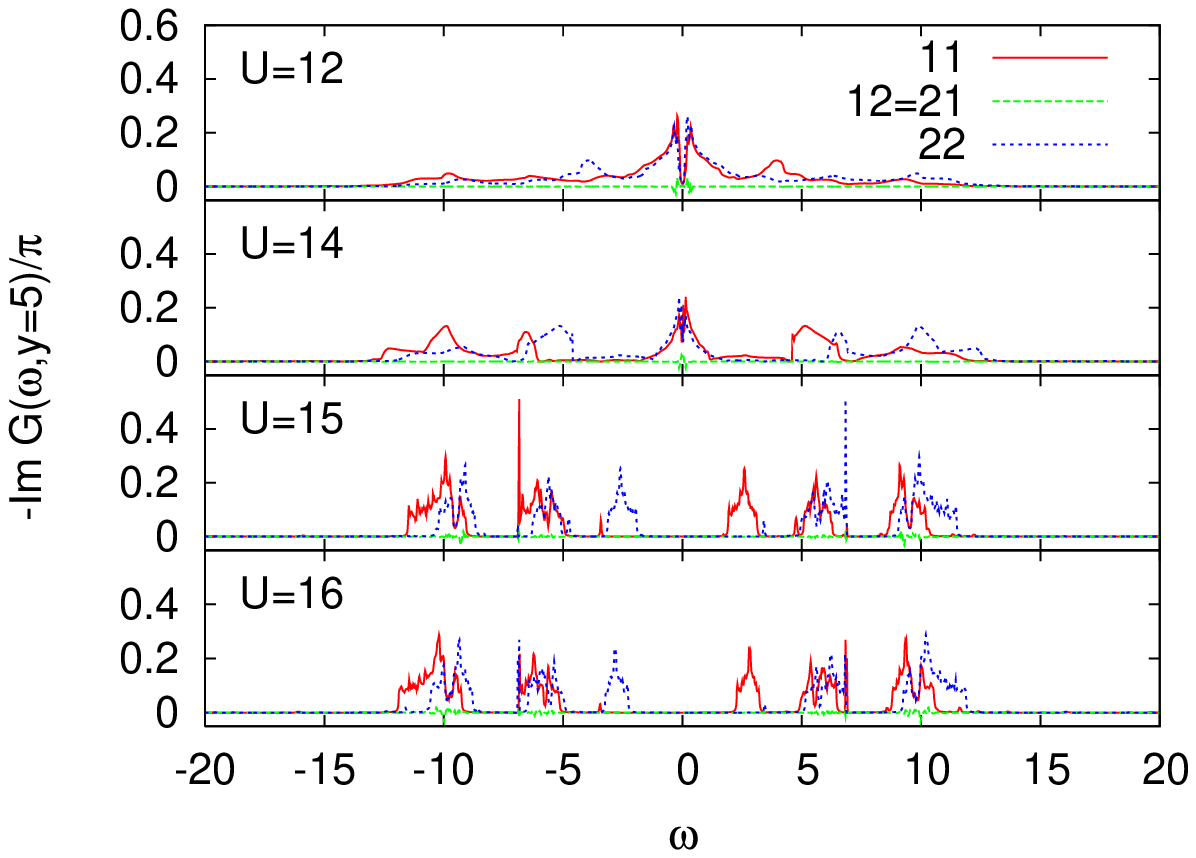}
\caption{(Color online) The imaginary parts of the local Green's functions
for several values of $U$ at $y=1$ (upper panel) and $y=5$ (lower panel)
when $L_y=10$.
Red, green, blue curves are $G_{11}, G_{12,21}$ and
$G_{22}$, respectively. 
}
\label{fig:ImG}
\end{figure}

\begin{figure}[tb]%
\includegraphics[width=\linewidth]{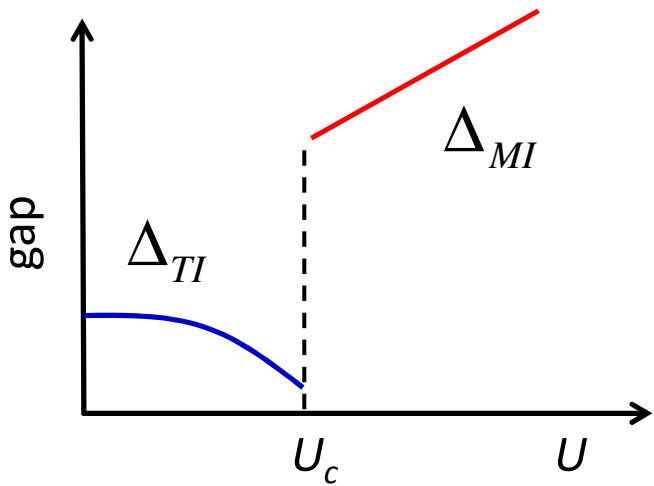}
\caption{(Color online) A schematic picture of the $U$-dependence of
$\Delta_{\rm TI}$ and $\Delta_{\rm MI}$.
At finite temperatures or if spatial correlations are taken
into account, $\Delta_{\rm TI}$ would not close
and the transition would be discontinuous.
}
\label{fig:schematic}
\end{figure}

Finally, we briefly discuss the gap structures in the imaginary part
of the local 
Green's functions which correspond to
the local density of states.
Figure \ref{fig:ImG} shows ${\rm Im}G$ for several different
interaction strengths $U$.
With increasing $U$ for $U<U_c$, the structures around the Fermi
energy become more and more renormalized, 
and finally for $U>U_c$ a
Mott gap with $\Delta_{\rm MI}\sim U$ opens.
Here, we note that the Mott transition as a topological
phase transition is qualitatively
different from transitions
in which parameters in the non-interacting part of the Hamiltonian are
changed.
For example, if the parameter $M_0$ is changed (for $U=0$), a
transition from a TI 
to a trivial band 
insulator will occur. However, the gap in both phases is well defined
in momentum-space.  
The gap width in each phase is changing continuously near the
topological phase transition. 
At the Mott transition, however, the well-defined gap in
momentum-space $\Delta_{\rm TI}$ for the TI 
phase is replaced by
the Mott gap $\Delta_{\rm MI}$
which arises due to a gap in the atomic 
electron configurations in real space.
Although the Mott transition would be continuous 
in the present calculations with the DMFT at zero temperature,
at finite temperatures or if spatial correlations are taken into account, 
the transition can be expected to be discontinuous.\cite{Georges96, 
Maier05, Yoshida11}  
The schematic behavior of the gap is
shown in Fig. \ref{fig:schematic}.

Finally, let us make some comments on $\Sigma_{12,21}$.
As mentioned in the previous section, the local selfenergy
satisfies $\Sigma_{12,21}(\omega,y)=-\Sigma_{12,21}(\omega,y)^{\ast}$,
and especially, ${\rm Re}\Sigma_{12,21}(\omega=0,y)=0$ for all
$y$ sites.
However, 
${\rm Im}\Sigma_{12,21}(\omega=0,y)$ does not necessarily have to vanish
due to symmetry arguments.
Although our calculated $\Sigma_{12,21}$ is much smaller
than $\Sigma_{11,22}$ and could be safely neglected,
we have small but finite values of $\Sigma_{12,21}(\omega=0,y)$ near
the edge sites. 
This might be related to the inter-edge Umklapp scattering.
However, since $\Sigma_{12,21}(\omega)$ shows complicated
$\omega$-dependence within the NRG calculations,
it is difficult to evaluate $\Sigma_{12,21}(\omega=0)$ with
sufficient accuracy.

\section{Summary and discussion}
\label{sec:summary}
In this article we have discussed correlation effects in
two dimensional topological insulators.
In the first part, we investigated finite size effects in
relatively large systems.
The effective theory for those edges states 
is given by one-dimensional spin-$1/2$ fermions including
inter-edge correlated forward scattering.

The Tomonaga-Luttinger parameters are $K_{s}\geq 1$ for the spin sector
and $K_{c}\leq 1$ for the charge sector. 
They include the renormalization factor defined in the whole two
dimensional system
and might not be so much away from unity.
The analysis of the scaling dimensions of the perturbations
revealed that the tunneling between two edges
is the most relevant perturbation around the non-interacting point when 
the Fermi wavenumber of the edge states is $k_F\sim0$.
We evaluated the gap size induced by the tunneling for each sector
which might be observed in experiments.

For small systems, the edge states are massive Dirac 
fermions and there are no massless single particle excitations.
We discussed the correlation effects by using the inhomogeneous
DMFT with the IPT in the ribbon geometry.
We confirmed that
the naive expectation based on the Fermi-liquid picture of
the correlated TIs, that the amplitude of the finite size gap is
simply renormalized by the averaged renormalization factor, and the
localization length of the edge states is not affected by the
interaction, actually holds for weak interactions.

Finally, we discussed the Mott transition using
the inhomogeneous DMFT+NRG.
The $\omega$-dependence of the selfenergy is qualitatively
similar to that in the metallic systems.
We could not see 
characteristic features in the selfenergy 
which signal the edge Mott insulating states,
although inter-edge Umklapp scattering is taken into account within DMFT.
We showed that correlation effects are significantly stronger
near the edge sites due to the reduction of the coordination number.
Since the edge states are localized around the edges,
the relatively strong correlations largely affect them especially
near the Mott transition.
As the interaction is tuned, we found a trivial Mott transition
from the topological insulating state to a nearly homogeneous
Mott insulating state.
The numerical results can also be understood within the linearized DMFT
as a result of the quantum tunneling of the electrons as has been
discussed for systems without bulk gaps.
The transition to a Mott insulator is characterized as a topological
phase transition, and  
the character of the gap in each phase is totally different
and the gap closing is not required at the transition in general.

In the present DMFT calculations, 
we have focused on the paramagnetic
phase and neglected instabilities to magnetically ordered 
phases.\cite{Yoshida11}
However, in addition to homogeneous magnetic states,
magnetic moments only near the edge sites might be
possible.\cite{Fujita96,DHLee11}
We also assumed that all parameters in the Hamiltonian are uniform,
although there can be changes especially near the edge sites in experimental
situations.
These issues are future problems.

\section*{Acknowledgment}
We thank T. Yoshida
for valuable discussions.
Numerical calculations were partially performed at the 
Yukawa Institute Computer Facility. 
This work is partly supported by the Grant-in-Aids for
Scientific Research from MEXT of Japan
(Grant No.19052003, 
Grant No.20102008, 
Grant No.20740194, 
Grant No.21102510, 
Grant No.21540359, 
Grant No.23540406, 
and Grant No.23840009), 
and the Grant-in-Aid for the Global COE Program 
"The Next Generation of Physics, Spun from Universality and Emergence"
and "Nanoscience and Quantum Physics".
RP thanks the Japan Society for the Promotion of Science (JSPS)
and the Alexander von Humboldt-Foundation.
NK is partly supported by JSPS through the "Funding Program for
World-Leading Innovative R\&D on Science and Technology (FIRST
Program)".

\appendix
\section{Details of NRG}
In this Appendix, we show details of the NRG formalism.
DMFT + NRG calculations with off-diagonal selfenergy 
have been performed in the context of
superconductivity,\cite{Bauer09,Bauer09PRB}
and our treatment is an extension of the
previous works.
We also discuss details of the numerical calculations.

\subsection{Effective Impurity Anderson model}
\label{chp:imp}
The impurity model arising from the DMFT self-consistency reads
\begin{eqnarray*}
H_{\rm IAM}(y)&=&H_c(y)+H_{hyb}(y)+H_{imp}(y),
\label{eq:A_IAM}\\
H_c&=&\sum_{ms,ll^{\prime}\sigma}a^{\dagger}_{l\sigma}(ms)
g_{ll^{\prime}\sigma}(ms)a_{l^{\prime}\sigma}(ms),\\
H_{hyb}&=&\sum_{ms,ll^{\prime}\sigma}
\Bigl[f_{l\sigma}^{\dagger}h_{ll^{\prime}\sigma}(ms)
a_{l^{\prime}\sigma}(ms)\\
&&+a^{\dagger}_{l^{\prime}\sigma}(ms)h_{ll^{\prime}\sigma}^{\ast}(ms)
f_{l\sigma}\Bigr],\\
H_{imp}&=&\sum_{l\sigma}E^f_{l\sigma}
f_{l\sigma}^{\dagger}f_{l\sigma}
+H_{int}[f_{l\sigma}^{\dagger},f_{l\sigma}]
\end{eqnarray*}
where 
\begin{displaymath}
E^f=(M_0-\mu,-M_0-\mu)\,.
\end{displaymath}
The impurity is represented by the operators $f_{l\sigma}$, where $l$
is the orbital- and $\sigma$ the spin index. The conduction electrons
are given by the operators $a_{l\sigma}(ms)$, where $ms$ corresponds
to an discretization interval $(m+)$, $(m-)$ as defined
below. Therefore, in this context $g_{ll\prime}$ corresponds to the
kinetic energy, while $h_{ll\prime}$ corresponds to the hybridization
between impurity and conduction electrons.
Note that all the parameters in $H_{\rm IAM}$ depend on the site index $y$,
which is suppressed for clarity of the notation.
From now on, we will neglect the spin indices, because the lattice 
Hamiltonian is diagonal in spin space.
The energy is discretized into intervals
\begin{eqnarray*}
I_{m+}&=&(x_{m+1},x_m]\\
I_{m-}&=&(-x_m,-x_{m+1}]\,,
\end{eqnarray*}
with $x_m=x_0\Lambda^{-m}$ ($m\in\mathbb{N}$).

We consider the particle-hole symmetric case and
assume that $g_{ll^{\prime}}(ms)\in\mathbb{R}$ and can be written as
\begin{eqnarray}
g(ms):=\left[
\begin{array}{cc}
\xi(ms) & {\mathcal L}(ms) \\
{\mathcal L}(ms) & -\xi(ms)
\end{array}
\right].
\label{eq:app_cond}
\end{eqnarray}
This assumption is legitimate, because our lattice Hamiltonian
is defined only by real values when using open boundary conditions
for the $y$-direction.
The hybridization $h$ is parametrized as
\begin{eqnarray*}
h(ms)=\left[
\begin{array}{cc}
a(ms) & b(ms) \\
b^{\prime}(ms) & c(ms)
\end{array}
\right],
\end{eqnarray*}
where $a, b, b^{\prime}$ and $c$ are also assumed to be real.

The kinetic energy term $g$ is diagonalized by the following
orthogonal matrices,
\begin{eqnarray*}
U(ms)&=&\left[
\begin{array}{cc}
u(ms) & -v(ms) \\
v(ms) & u(ms)
\end{array}\right],\\
u^2(ms)&=&\frac{1}{2}\left(1+\frac{\xi(ms)}{E(ms)}\right),\\
v^2(ms)&=&\frac{1}{2}\left(1-\frac{\xi(ms)}{E(ms)}\right),\\
E(ms)&=&\sqrt{\xi^2(ms)+{\mathcal L}^2(ms)}.
\end{eqnarray*}
$E(ms)$ can finally be identified (for both $s=\pm$) as
\begin{eqnarray*}
E(ms)=|x_m+x_{m+1}|/2,
\end{eqnarray*}
corresponding to the discretized intervals.
The kinetic energy and the hybridization are transformed as
\begin{eqnarray*}
\tilde{g}(ms)&=&U^{\dagger}g(ms)U\\
&=&\left[
\begin{array}{cc}
E(ms) & 0 \\
0 & -E(ms)
\end{array}
\right],\\
\tilde{h}(ms)&=&h(ms)U(ms)\\
&=&\left[
\begin{array}{cc}
au+bv & -av+bu \\
b^{\prime}u+cv & -b^{\prime}v+cu
\end{array}
\right]\\
&\equiv&\left[
\begin{array}{cc}
A & B \\
B^{\prime} & C
\end{array}
\right].
\end{eqnarray*}
Then, in this new basis $\tilde{a}$, the impurity Anderson
Hamiltonian is written as
\begin{eqnarray*}
H_c&=&\sum_{ms,\alpha}\tilde{a}^{\dagger}_{\alpha}(ms)
E_{\alpha}(ms)\tilde{a}_{\alpha}(ms),\\
H_{hyb}&=&\!\!\!\sum_{ms,l,\alpha}[f_{l}^{\dagger}\tilde{h}_{l\alpha}(ms)
\tilde{a}_{\alpha}(ms)
+\tilde{a}^{\dagger}_{\alpha}(ms)\tilde{h}_{l\alpha}(ms)
f_{l}],\\
\tilde{a}_{\alpha}(ms)&=&\sum_{l}U^{\dagger}_{\alpha l}(ms)
a_{l}(ms),
\end{eqnarray*}
where $E_{\alpha}(ms)=(E(ms),-E(ms))$.

The retarded non-interacting Green's function of the impurity
Anderson model can be
written as,
\begin{eqnarray*}
{\mathcal G}^{-1}(\omega+i0)=\omega-E^f-K(\omega+i0),
\end{eqnarray*}
where the hybridization $K$ is given as,
\begin{eqnarray*}
K_{ll^{\prime}}(\omega)&=&
\sum_{ms}\left(h(ms)[\omega-g(ms)]^{-1}h^{\dagger}(ms)
\right)_{ll^{\prime}}\\
&=&\sum_{ms}\left((hU)[\omega-U^{\dagger}gU]^{-1}(U^{\dagger}h^{\dagger})
\right)_{ll^{\prime}}\\
&=&\sum_{ms}\sum_{\alpha}
\frac{\tilde{h}_{l\alpha}\tilde{h}_{l^{\prime}\alpha}}
{\omega-E_{\alpha}}.
\end{eqnarray*}
Using the identity 
$1=\int d\varepsilon \delta(\varepsilon-E_{\alpha}(ms))$ 
for each $\alpha$,
we obtain
\begin{eqnarray*}
K_{ll^{\prime}}(\omega)&=&
\int d\varepsilon \frac{\Delta_{ll^{\prime}}(\varepsilon)}{\omega
-\varepsilon+i0},\\
\Delta_{ll^{\prime}}(\varepsilon)&=&\sum_{ms,\alpha}
\tilde{h}_{l\alpha}(ms)\tilde{h}_{l^{\prime}\alpha}(ms)
\delta(\varepsilon-E_{\alpha}(ms)).
\end{eqnarray*}
We note that due to the Kramers-Kronig relation,
\begin{eqnarray*}
\Delta(\omega)=-\frac{1}{\pi}{\rm Im}K(\omega)
\end{eqnarray*}
is satisfied.

Next, for each interval $I_{ms}$, we define
\begin{eqnarray*}
\Delta(ms)=\int _{I_{ms}}d\omega \Delta(\omega).
\end{eqnarray*}
Thus, for the integral region $I_{m_0+}$
which includes only a single value $E(ms)\in
\{E(n\sigma)\}_{n\sigma}$, the following equations hold
\begin{eqnarray}
\Delta_{ll^{\prime}}(m+)=\sum_s
\tilde{h}_{l 1}(ms)\tilde{h}_{l^{\prime} 1}(ms),
\label{eq:cond_p}
\end{eqnarray}
and for $I_{m_0-}$,
\begin{eqnarray}
\Delta_{ll^{\prime}}(m-)=\sum_s
\tilde{h}_{l 2}(ms)\tilde{h}_{l^{\prime} 2}(ms).
\label{eq:cond_m}
\end{eqnarray}

The number of obtained equations for $\tilde{h}$ 
given by Eqs. (\ref{eq:cond_p}) and (\ref{eq:cond_m}) is six.
$\Delta_{11}(m\pm),\Delta_{22}(m\pm),\Delta_{12}(m\pm)$ are
involved and $\Delta_{21}$ gives the same equations as $\Delta_{12}$.
However, the number of independent $\tilde{h}_{l\alpha}$ is eight
and two extra conditions are needed for $\tilde{h}_{l\alpha}$
to be determined.
One possibility is that
extra conditions are imposed so that 
$\tilde{h}$ is connected smoothly to the trivial situation if 
we turn off the SO interaction,
which is legitimate because $|h_{11,22}|>|h_{12,21}|$ is expected
in general physical situations.

Taking the limit ${\mathcal L}\rightarrow +0$ for the entries of
the conduction electrons of the original
untransformed impurity model (Eq \ref{eq:app_cond})
and because $\xi(m+)>0$ and $\xi(m-)<0$ can be assumed,
the limiting behaviors of $u,v$ should be
\begin{eqnarray*}
u(m+)\rightarrow 1,\quad v(m+)\rightarrow 0,\\
u(m-)\rightarrow 0,\quad v(m-)\rightarrow 1,
\end{eqnarray*}
and
\begin{eqnarray*}
\tilde{h}(m+)&\rightarrow& \left[
\begin{array}{cc}
a(m+) & b(m+) \\
b^{\prime}(m+) & c(m+)
\end{array}
\right],\\
\tilde{h}(m-)&\rightarrow& \left[
\begin{array}{cc}
b(m-) & -a(m-) \\
c(m-) & -b^{\prime}(m-)
\end{array}
\right].
\end{eqnarray*}
Therefore, a possible additional condition which keeps the symmetry of $\tilde{h}$
is, $A(m-)=-B^{\prime}(m+)$ and $C(m-)=+B(m+)$.
These conditions are linear in $\tilde{h}$.
In the limit ${\mathcal L}\rightarrow 0$,
the first equation corresponds to $b_-=-b_+^{\prime}$,
and the second one to $-b_-^{\prime}=b_+$.
Here, we have used shortened expressions $X_{s}=Y_{s^{\prime}}$ for
$X(ms)=Y(ms^{\prime})$.
Then, the equations for $\tilde{h}$ are,
\begin{eqnarray*}
\Delta_{11+}&=&A_+^2+A_-^2,\\
\Delta_{22+}&=&B_+^{\prime 2}+B_-^{\prime 2}=A_-^2+B_-^{\prime 2},\\
\Delta_{12+}&=&A_+B_+^{\prime}+A_-B_-^{\prime}=A_-(-A_++B_-^{\prime}),\\
\Delta_{11-}&=&B_+^2+B_-^2=C_-^2+B_-^2,\\
\Delta_{22-}&=&C_+^2+C_-^2,\\
\Delta_{12-}&=&C_+B_++C_-B_-=C_-(C_++B_-)
\end{eqnarray*}
In this set of equations the upper three and lower three equations
are decoupled. 
Because these equations are non-linear and have several
possible solutions, the solution
which can be smoothly connected to the limit ${\mathcal L}\rightarrow +0$
and $b,b^{\prime}\rightarrow 0$ should be carefully chosen.
The results are,
\begin{eqnarray*}
\delta_s&\equiv&-\frac{(\Delta_{12s})^2}
{(\Delta_{11s}-\Delta_{22s})^2+4\Delta_{12s}^2}
\Bigl[(\Delta_{11s}+\Delta_{22s})\\&&- 
2\sqrt{\Delta_{11s}\Delta_{22s}-\Delta_{12s}^2}\Bigr],\\
A_+&=&\sqrt{\Delta_{11+}+\delta_+},\\
A_-&=&-\frac{\Delta_{12+}}{\sqrt{\Delta_{11+}+\delta_+}
+\sqrt{\Delta_{22+}+\delta_+}},\\
B_+^{\prime}&=&-A_-,\\
B_-^{\prime}&=&-\sqrt{\Delta_{22+}+\delta_+},\\
C_+&=&-\sqrt{\Delta_{22-}+\delta_-},\\
C_-&=&-\frac{\Delta_{12-}}{\sqrt{\Delta_{11-}+\delta_-}
+\sqrt{\Delta_{22-}+\delta_-}},\\
B_+&=&C_-,\\
B_-&=&-\sqrt{\Delta_{11-}+\delta_-}.
\end{eqnarray*}
Note that for the above solution of $\delta_{s}$,
\begin{eqnarray*}
\lim_{\Delta_{22s}\rightarrow \Delta_{11s}}
\lim_{\Delta_{12s}\rightarrow 0}\delta_s
=\lim_{\Delta_{12s}\rightarrow 0}
\lim_{\Delta_{22s}\rightarrow \Delta_{11s}}\delta_s
=0
\end{eqnarray*}
is satisfied and $\delta_s$ is smoothly connected to the case without the SO
interaction.
Because we focus on the particle-hole symmetric
case with $\Delta_{11\pm}=\Delta_{22\mp},\Delta_{12+}
=\Delta_{12-}$, some relations can be derived from the above expressions of 
$A\sim C$:
$\delta_+=\delta_-,A_+=-C_+,B_+^{\prime}=-B_+,A_-=C_-$
and $B_-^{\prime}=B_-$.
The resulting $\tilde{h}$ has the form of
\begin{eqnarray*}
\tilde{h}(m+)&=&\left[
\begin{array}{cc}
A(m+) & B(m+) \\
-B(m+) & -A(m+)
\end{array}
\right]\\
&\rightarrow &
\left[
\begin{array}{cc}
\sqrt{\Delta_{11}(m+)} & 0 \\
0 & -\sqrt{\Delta_{11}(m+)}
\end{array}
\right],\\
\tilde{h}(m-)&=&\left[
\begin{array}{cc}
A(m-) & B(m-) \\
B(m-) & A(m-)
\end{array}
\right]\\
&\rightarrow &
\left[
\begin{array}{cc}
0 & -\sqrt{\Delta_{11}(m-)} \\
-\sqrt{\Delta_{11}(m-)} & 0
\end{array}
\right],
\end{eqnarray*}
where '$\rightarrow$' means the limit 
${\mathcal L}\rightarrow 0$ and $b,b^{\prime}\rightarrow 0$.
Correspondingly, $h(ms)=\tilde{h}(ms)U^{\dagger}(ms)\rightarrow
{\rm diag}(\sqrt{\Delta_{11}(ms)},-\sqrt{\Delta_{11}(ms)})$.

We note that a condition is required for $A,B,B^{\prime}$ and $C$ to be
real quantities:
\begin{eqnarray}
-{\rm max}\{ \Delta_{11s},\Delta_{22s}\}\leq \delta_s\leq 0.
\end{eqnarray}
Another condition is needed for $\delta_{\pm}$ to be real:
\begin{eqnarray}
\Delta_{11s}\Delta_{22s}-\Delta_{12s}^2\geq 0.
\end{eqnarray}
These two conditions are essentially equivalent.
The above derivation is based on the assumption that
$\tilde{h}$ should be connected smoothly to the one without
the SO interaction, ${\mathcal L},b,b^{\prime}\rightarrow 0$.
If one assumes other smooth connections, such as a
connection to $\xi \rightarrow 0$, different expressions
for $A\sim C$ would be obtained.

\subsection{Chain Hamiltonian}
\label{app:chain}
In the next step we transform $H_{\rm IAM}$ into
a chain Hamiltonian.
We want to transform the impurity Anderson model to
\begin{eqnarray*}
H_c&=&\sum_{n\nu \nu^{\prime} \sigma}\varepsilon_{n\nu \nu^{\prime}}
f^{\dagger}_{n\nu \sigma}f_{n\nu^{\prime} \sigma}\\
&+&\sum_{n\nu \nu^{\prime}\sigma}
t_{n\nu \nu^{\prime}}[f^{\dagger}_{n\nu \sigma}f_{n+1\nu^{\prime} \sigma}
+f^{\dagger}_{n+1\nu^{\prime} \sigma}f_{n\nu \sigma}],\\
H_{hyb}&=&\sum_{\nu \nu^{\prime} \sigma}v_{\nu \nu^{\prime}}
[f^{\dagger}_{-1\nu \sigma}f_{0\nu^{\prime} \sigma}+
f^{\dagger}_{0\nu^{\prime} \sigma}f_{-1\nu \sigma}],\\
H_{imp}&=&\sum_{\nu \sigma}E^f_{\nu}f^{\dagger}_{-1\nu \sigma}f_{-1\nu \sigma}
+H_{int}[f^{\dagger}_{-1\nu \sigma},f_{-1\nu \sigma}].
\end{eqnarray*}
Here,
$f_{-1\nu}$ corresponds to the former $f_{\alpha}$,
\begin{eqnarray*}
f_{-1\nu}&=&\sum_{\alpha}\delta_{\alpha \nu}f_{\alpha}.
\end{eqnarray*}
This chain Hamiltonian is schematically shown in Fig. \ref{fig:IAM} of
the main text.
Because the impurity part $H_{imp}$ is not transformed from its
original expression and $H_c$ again includes the inter-chain hopping,
we understand that the index $\nu$ in the basis $\{f_{n\nu\sigma}\}$
characterizes the orbitals $l=1, 2$.
As far as we know, impurity Anderson models 
with inter-chain hopping have never
been studied by using NRG.
The inter-chain hopping in the present model arises from the 
inter-orbital hopping, and it makes numerical calculations hard.

Let us assume that this transformation is
given by an orthogonal matrix $u=(u_{nms}^{\nu a})$,
\begin{eqnarray*}
f_{n\nu}&=&\sum_{msa}u_{nms}^{\nu\alpha}\tilde{a}_{\alpha}(ms),\\
\tilde{a}_{\alpha}(ms)&=&\sum_{n\nu}u_{nms}^{\nu\alpha}f_{n\nu},\\
\sum_{ms\alpha}u_{nms}^{\nu\alpha}u_{n^{\prime}ms}^{\nu^{\prime}\alpha}&=&
\delta_{nn^{\prime}}\delta_{\nu \nu^{\prime}},\\
\sum_{n\nu}u_{nms}^{\nu\alpha}u_{nm^{\prime}s^{\prime}}^{\nu\alpha^{\prime}}
&=&\delta_{mm^{\prime}}\delta_{ss^{\prime}}\delta_{\alpha \alpha^{\prime}}.
\end{eqnarray*}
Then, the hybridization term would be,
\begin{eqnarray*}
H_{hyb}&=&\sum_{msl\alpha}[f^{\dagger}_{l}\tilde{h}_{l\alpha}(ms)
\tilde{a}_{\alpha}(ms)+{\rm h.c.}]\\
&=&\sum_{n\nu \nu^{\prime}}f^{\dagger}_{-1\nu}\left[ 
\sum_{ms\alpha}\tilde{h}_{\nu\alpha}(ms)u_{nms}^{\nu^{\prime}\alpha}
\right]f_{n\nu^{\prime}}
+({\rm h.c.}).
\end{eqnarray*}
We define a vector $(\tilde{u}^{\nu}_0)^{\alpha}_{ms}=
\tilde{u}^{\nu\alpha}_{0ms}$ and an inner product as
\begin{eqnarray*}
\tilde{u}^{\nu\alpha}_{0ms}&=&\tilde{h}_{\nu\alpha}(ms),\\
\langle u_n^{\nu},u_{n^{\prime}}^{\nu^{\prime}}\rangle
&\equiv&
\sum_{ms\alpha}u^{\nu\alpha}_{nms}u_{n^{\prime}ms}^{\nu^{\prime}\alpha},
\end{eqnarray*}
and normalize $\tilde{u}^{\nu\alpha}_{0ms}$ as 
$\bar{u}^{\nu\alpha}_{0ms}=\tilde{u}^{\nu\alpha}_{0ms}/
\sqrt{|\langle \tilde{u}_0^{\nu},u_0^{\nu^{\prime}}\rangle|}$.
We, then, orthogonalize $\{\bar{u}_0^{1},\bar{u}_0^2\}$ keeping their
symmetry to obtain $\{u_0^{1},u_0^2\}$.
To fulfill symmetry requirements, the trigonalization must be
performed always by including two vectors.
Therefore, we assume that orthogonalization can be performed by
\begin{eqnarray*}
\bar{u}_0^{1new}&=&\bar{u}^1_0+c\langle \bar{u}_0^{1},
\bar{u}_0^{2}\rangle \bar{u}_0^{2},\\
\bar{u}_0^{2new}&=&\bar{u}^2_0+c\langle \bar{u}_0^{1},
\bar{u}_0^{2}\rangle \bar{u}_0^{1}
\end{eqnarray*}
with one coefficient $c$.
Then, $\langle \bar{u}_0^{1new},\bar{u}_0^{2new}\rangle
=\langle \bar{u}_0^{1},\bar{u}_0^{2}\rangle
[1+2c+c^2\langle \bar{u}_0^{1},\bar{u}_0^{2}\rangle]$.
If $c=-0.5$, $\langle \bar{u}_0^{1new},\bar{u}_0^{2new}\rangle=
0.25\times \langle \bar{u}_0^{1},\bar{u}_0^{2}\rangle ^2$.
Because $\bar{u}_0^1$ and $\bar{u}_0^2$ have been normalized, 
$|\langle \bar{u}_0^{1},\bar{u}_0^{2}\rangle |<1$
is satisfied. Repeating this procedure
leads to orthogonalized unit vectors $\{u_0^{1},u_0^2\}$.
The hopping from site $-1$ to site $0$ is calculated
from $\tilde{u}_{0ms}^{\nu\alpha}=\tilde{h}_{\nu\alpha}(ms)$ as
\begin{eqnarray*}
v_{\nu \nu^{\prime}}=
\langle \tilde{u}_0^{\nu},u_0^{\nu^{\prime}}\rangle.
\end{eqnarray*}

Next, let us move to the kinetic term.
\begin{eqnarray*}
H_{c}&=&\sum_{ms\alpha}E_{\alpha}(ms)\tilde{a}_{\alpha}^{\dagger}(ms)
\tilde{a}_{\alpha}(ms)\\
&=&\sum_{n\nu}\sum_{n^{\prime}\nu^{\prime}}
\left[ \sum_{ms\alpha}E_{\alpha}(ms)u^{\nu\alpha}_{nms}
u^{\nu^{\prime}\alpha}_{n^{\prime}ms}\right]
f^{\dagger}_{n\nu}f_{n^{\prime}\nu^{\prime}}.
\end{eqnarray*}
This claims that 
\begin{eqnarray*}
\sum_{ms\alpha}E_{\alpha}(ms)u^{\nu \alpha}_{nms}
u^{\nu^{\prime}\alpha}_{n^{\prime}ms}
&=&\varepsilon_{n\nu \nu^{\prime}}\delta_{nn^{\prime}}
\\&+&t_{n\nu \nu^{\prime}}[\delta_{n^{\prime},n+1}+\delta_{n,n^{\prime}+1}]
\end{eqnarray*}
should hold.

If $u_{j}^{\nu},j=0,\cdots ,n$ are obtained, 
we calculate $\varepsilon_{n\nu \nu^{\prime}}$ from
\begin{eqnarray*}
\varepsilon_{n\nu \nu^{\prime}}
=\sum_{ms\alpha}E_{\alpha}(ms)u^{\nu\alpha}_{nms}u^{\nu^{\prime}\alpha}_{nms}.
\end{eqnarray*}
Then, we define
\begin{eqnarray*}
\tilde{u}^{\nu\alpha}_{n+1ms}&=&\sum_{\nu^{\prime},ms\alpha}
\Bigl[(E_{\alpha}(ms)\delta_{\nu \nu^{\prime}}-\varepsilon_{n\nu \nu^{\prime}})
u_{nms}^{\nu^{\prime}\alpha}\\&&-t_{n-1\nu \nu^{\prime}}
u_{n-1ms}^{\nu^{\prime}\alpha}\Bigr],
\end{eqnarray*}
and normalize each of $\tilde{u}^{1,2}_{n+1}$ to obtain
normalized vectors $\bar{u}^{1,2}_{n+1}$.
Using $\bar{u}^{1,2}_{n+1}$ as initial vectors,
we perform the orthogonalization introduced above.
and obtain orthogonal vectors $\{ u^{1}_{n+1},u^{2}_{n+1}\}$.
Usual Schmidt orthogonalization 
should also be imposed
so that $\langle u_{n+1}^{\nu},u_j^{\nu^{\prime}}\rangle=0$
for $j=0,\cdots ,n$ are satisfied.
Then, we evaluate $t_{n}$ as
\begin{eqnarray*}
t_{n\nu \nu^{\prime}}
=\sum_{ms\alpha}E_{\alpha}(ms)u^{\nu\alpha}_{nms}
u^{\nu^{\prime}\alpha}_{n+1ms}.
\end{eqnarray*}
All the parameters in the above decrease exponentially, and
the input $\Delta$ is well reproduced from the chain Hamiltonian
derived in this way.

We note that, although our chain impurity Anderson model is rather complicated,
it is smoothly connected to the one without the inter-orbital hopping which
is nothing but two decoupled single-orbital impurity Anderson models 
when there are
no inter-orbital interactions at the impurity site.
The ground state of our impurity Anderson model 
is simply the Kondo screened state or
the decoupled state depending on the structure of the hybridization
$\Delta$.

\subsection{Discretization, Truncation and Broadening}
\label{ch:Disc}
\begin{figure}[tb]
\includegraphics[width=\linewidth]{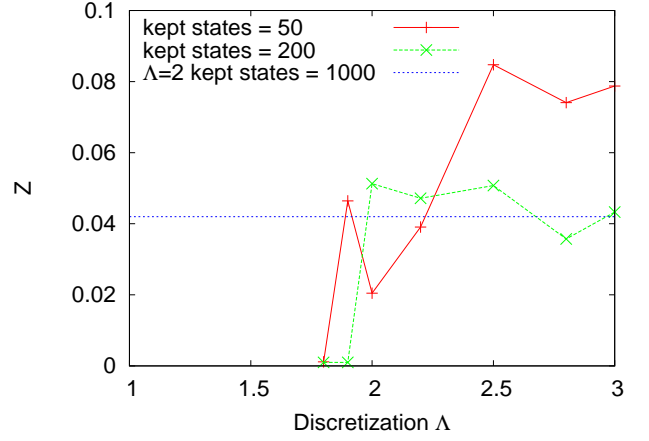}
\caption{(Color online) Truncation and discretization effects for the
  metal insulator transition in the ordinary one-orbital Hubbard
  model \cite{bulla1999}. We show the renormalization $z$ for
$U=1.4W\approx 0.94U_c$.
  The lines are meant as guide to the eye. The blue
  line is the reference value calculated for $\Lambda=2$,
  $N_K=1000$ (independent of the discretization parameter).
\label{trunc_dis}}
\end{figure}
As already mentioned in the main text, the NRG is not able to gain its
usual accuracy around the Fermi energy due to the
spin-orbit interaction which creates an hybridization between the two
orbitals. For the most of our
shown results we used a discretization $\Lambda=3.0$ and a truncation
of $N_K=5000$ states. Although $N_K=5000$ may sound a large
number of kept states, one should compare this number to that for a single
impurity Anderson model (SIAM) for which similar accuracy with
$N_{SIAM}=\sqrt{5000}\approx 70$ states would be obtained. Reducing
the discretization parameter $\Lambda$ to the often used value of
$\Lambda=2.0$ results in a breaking of the
symmetries discussed above, e.g.
($G_{12}(\omega)\neq G_{21}(\omega)$) which
are supposed to be satisfied. To reduce the truncation effects and
restore the symmetries we had to use a discretization parameter
$\Lambda=3.0$.

Therefore we briefly want to discuss the effects of discretization and
truncations for the well studied metal insulator transition in the
one-orbital Hubbard model at half filling.\cite{bulla1999} In 
Fig. \ref{trunc_dis} we show 
the z-values 
$z=[1-\partial \Sigma/\partial \omega]^{-1}_{\omega=0}$ for
different truncations and discretization parameters $\Lambda$ for an
interaction strength $U=1.4W\approx 0.94U_c$
close to the
Mott transition at $U_c$.
Large
discretization parameters $\Lambda$ lead to faster decreasing
hopping parameters along the Wilson chain, reducing the effects of 
truncated high-energy states on the kept low-energy states, thus
restoring the fundamental working principle of NRG. On the other hand, a large
discretization parameter reduces the available number of states, 
and calculation results go
away from those for the continuum limit $\Lambda\rightarrow1$. 

In the usual NRG iteration scheme, the renormalization procedure
starts at high energies 
(large frequencies in the spectral functions) and iterates down to the
Fermi energy (zero frequency in the spectral functions).
Thus, using small discretization parameter $\Lambda$ and small
number of kept states, 
the states that would actually couple and change the low energy
spectrum, are truncated at high energies and not correctly taken into
account.

From Fig. \ref{trunc_dis}, one can read off a clear tendency for the 
calculations including strong truncation effects: correlation effects
are overestimated in calculations using
small discretization parameters 
$\Lambda$, and possibily, small $\Lambda$ even
lead to an insulating solution. Large discretization parameters, on
the other hand, seem to underestimate the correlation effects. 

\begin{figure}[tb]
\includegraphics[width=\linewidth]{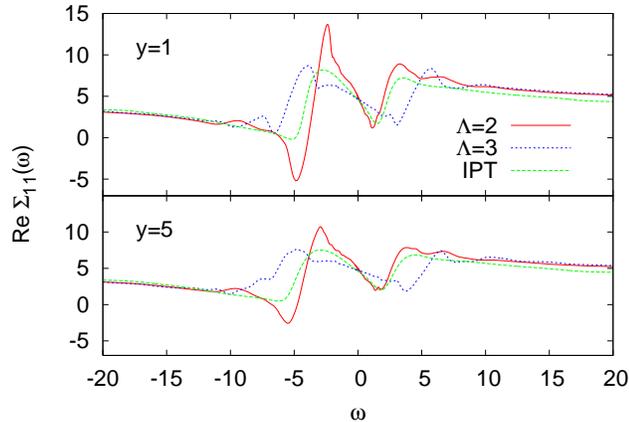}
\caption{
{(Color online) 
Real part of the self energy for $U=8$ between two discretization
  parameters in NRG and IPT as impurity solver. The upper panel shows
  y-site $1$, the lower panel y-site $5$.}
\label{comparison_Lambda}
}
\end{figure}
For checking the presented results we have performed some additional test
calculations reducing the discretization parameter $\Lambda$ but
increasing the number of kept states. Figure \ref{comparison_Lambda}
shows a comparison for the real part of the selfenergy for
($\Lambda=3$, $N_K=5000$), ($\Lambda=2$, $N_K=8000$), and IPT as an
impurity solver. 
The tendency agrees with the results for the
one-orbital Hubbard model suffering from the large truncation effects, see
Fig. \ref{trunc_dis}. The correlation effects for $\Lambda=2$ are much
more pronounced than for $\Lambda=3$. The IPT result lies between
the two NRG results. Furthermore, increasing the interaction strength using
discretization $\Lambda=2$, leads to a Mott transition at $U\approx
12$. However, in all the calculations using $\Lambda=2$ we find no signs
of edge Mott insulating states and all claims made in the main text
remain valid. 
\bibliography{submit}


\end{document}